\documentclass[12pt]{article}
\pdfoutput=1
\usepackage{amsmath,amsfonts,amssymb,amsthm,amssymb,bbm,bm}
\usepackage{pdfsync}
\usepackage{graphicx,import}
\usepackage[utf8]{inputenc}
\usepackage{empheq}
\usepackage[all]{xy}
\usepackage{stmaryrd}
\usepackage{rotating}
\usepackage[dvipsnames]{xcolor}  
\usepackage{slashed}
\usepackage{cancel}
\usepackage{array, makecell} %
\usepackage{comment}
\usepackage{tikz-cd}
\usepackage{hhline}
\usepackage{cancel}
\usepackage{dsfont}
\usepackage{stackengine}
\usepackage[breakable]{tcolorbox}
\usepackage{mathrsfs}
\usepackage{changepage}

\allowdisplaybreaks

\usepackage[pagebackref]{hyperref}
\renewcommand*{\backref}[1]{}
\renewcommand*{\backrefalt}[4]{\small{
    \ifcase #1%
          \or (Cited on page~#2.)%
          \else (Cited on pages~#2.)%
    \fi%
    }}
\hypersetup{colorlinks,linkcolor={RedViolet},citecolor={Blue},urlcolor={Blue}}

\usepackage{orcidlink}

\DeclareMathAlphabet\mathrsfso{U}{rsfso}{m}{n}

\usepackage[scr=boondoxo,scrscaled=1.1]{mathalpha}

\makeatletter

\DeclareFontFamily{U}{matha}{\hyphenchar\font45}
\DeclareFontShape{U}{matha}{m}{n}{
      <5> <6> <7> <8> <9> <10> gen * matha
      <10.95> matha10 <12> <14.4> <17.28> <20.74> <24.88> matha12
      }{}
\DeclareSymbolFont{matha}{U}{matha}{m}{n}
\DeclareMathSymbol{\oright}       {2}{matha}{"69}

\usepackage[top=2cm, bottom=2.5cm, left=2.5cm, right=2.5cm]{geometry}

\newcommand{\doublehat}[1]{%
\begingroup%
  \let\macc@kerna\z@%
  \let\macc@kernb\z@%
  \let\macc@nucleus\@empty%
  \hat{\raisebox{.55ex}{\vphantom{\ensuremath{#1}}}\smash{\hat{#1}}}%
\endgroup%
}

\renewcommand{\ni}{\noindent}

\newcommand{\bit}{\begin{itemize}}
\newcommand{\eit}{\end{itemize}}
\newcommand{\bd}{\begin{description}}
\newcommand{\ed}{\end{description}}

\newcommand{\bc}{\begin{center}}
\newcommand{\ec}{\end{center}}

\newcommand{\N}{{\mathbb N}}
\newcommand{\R}{{\mathbb R}}
\newcommand{\E}{\mathsf{E}}

\newcommand{\cA}{{\mathcal A}}

\def\be#1\ee{\begin{align}#1\end{align}}
\newcommand{\bea}{\begin{eqnarray}}
\newcommand{\eea}{\end{eqnarray}}
\newcommand{\bs}{\begin{subequations}}
\newcommand{\es}{\end{subequations}}
\newcommand{\nn}{\nonumber}

\usepackage[numbers,sort&compress]{natbib}

\makeatletter
\newcommand{\oset}[3][0ex]{%
  \mathrel{\mathop{#3}\limits^{
    \vbox to#1{\kern-2\ex@
    \hbox{$\scriptstyle#2$}}}}}
\makeatother

\def\rd{\mathrm{d}}
\def\pa{\partial}
\newcommand{\Dcal}{\mathcal{D}}

\newcommand{\pui}[1][1]{\pa_u^{-#1}}

\newcommand{\da}{\delta_\alpha}
\newcommand{\dap}{\delta_{\alpha'}}
\newcommand{\dapp}{\delta_{\alpha''}}
\newcommand{\dah}{\delta_{\fa}}

\newcommand{\dT}{\delta_T}

\newcommand{\dA}{\delta_{\text{\scalebox{1.1}{$\mathsf{a}$}}}}
\newcommand{\dAp}{\delta_{\text{\scalebox{1.1}{$\mathsf{a}$}$'$}}}

\renewcommand{\d}[2]{\delta^{\scriptscriptstyle{[#1]}}_{#2}}
\newcommand{\td}[2]{\tilde{\delta}^{\scriptscriptstyle{[#1]}}_{#2}}

\newcommand{\Hd}[2]{\delta^{\scriptscriptstyle{[#1]}\mathsf{H}}_{#2}}
\newcommand{\Htd}[2]{\tilde{\delta}^{\scriptscriptstyle{[#1]}\mathsf{H}}_{#2}}

\newcommand{\Sd}[2]{\delta^{\scriptscriptstyle{[#1]}\mathsf{S}}_{#2}}
\newcommand{\Std}[2]{\tilde{\delta}^{\scriptscriptstyle{[#1]}\mathsf{S}}_{#2}}

\newcommand{\sE}{\mathsf{E}}

\newcommand{\overbar}[1]{\mkern 3mu\overline{\mkern-3.5mu#1\mkern-2mu}\mkern 2mu}

\newcommand{\overbarF}[1]{\mkern 2mu\overline{\mkern-4mu#1\mkern-1mu}\mkern 2mu}

\newcommand{\bnews}{\overbarF{F}}
\newcommand{\bA}{\bar{A}}

\newcommand{\bz}{\bar{z}}
\renewcommand{\bm}{\overbar{m}}
\newcommand{\bY}{\overbar{Y}}
\newcommand{\bD}{\overbar{D}}

\newcommand{\hA}{\widehat{A}}
\newcommand{\hD}{\widehat{\Delta}}

\newcommand{\hq}{\widehat{\cq}}

\newcommand{\fa}{\hat{\alpha}}

\makeatletter
\newcommand{\hhat}[1]{%
\begingroup%
  \let\macc@kerna\z@%
  \let\macc@kernb\z@%
  \let\macc@nucleus\@empty%
  \widehat{\mathchoice%
    {\raisebox{.2ex}{\vphantom{\ensuremath{\displaystyle #1}}}}%
    {\raisebox{.4ex}{\vphantom{\ensuremath{\textstyle #1}}}}%
    {\raisebox{.16ex}{\vphantom{\ensuremath{\scriptstyle #1}}}}%
    {\raisebox{.14ex}{\vphantom{\ensuremath{\scriptscriptstyle #1}}}}%
    \smash{\widehat{#1}}}%
\endgroup%
}
\makeatother

\makeatletter
\DeclareRobustCommand\widecheck[1]{{\mathpalette\@widecheck{#1}}}
\def\@widecheck#1#2{%
    \setbox\z@\hbox{\m@th$#1#2$}%
    \setbox\tw@\hbox{\m@th$#1%
       \widehat{%
          \vrule\@width\z@\@height\ht\z@
          \vrule\@height\z@\@width\wd\z@}$}%
    \dp\tw@-\ht\z@
    \@tempdima\ht\z@ \advance\@tempdima2\ht\tw@ \divide\@tempdima\thr@@
    \setbox\tw@\hbox{%
       \raise\@tempdima\hbox{\scalebox{1}[-1]{\lower\@tempdima\box
\tw@}}}%
    {\ooalign{\box\tw@ \cr \box\z@}}}
\makeatother

\newcommand{\tQ}{\widetilde{Q}}
\newcommand{\tq}{\mkern0.7mu\widetilde{\cq}}

\newcommand{\Tr}{\mathrm{Tr}}
\newcommand{\news}{F}

\newcommand{\gYM}{g_\textsc{ym}}

\newcommand{\X}{\mathfrak{X}}
\newcommand{\g}{\mathfrak{g}}

\newcommand{\G}[1]{\mathcal{G}\big(#1\big)}

\newcommand{\PS}{\mathcal{P}}

\newcommand{\cS}{\mathcal{S}}
\newcommand{\WS}[1]{\mathcal{W}^{#1}_A(S)}

\newcommand{\V}{\mathsf{V}}
\newcommand{\tV}{\widetilde{\mathsf{V}}}

\newcommand{\W}{\mathsf{W}}
\newcommand{\Wcal}{\mathcal{W}}

\newcommand{\sfS}{\mathsf{S}}

\newcommand{\Ccel}[1]{\mathcal{C}^\textsf{cel}_{#1}(S)}
\newcommand{\Ccar}[1]{\mathcal{C}^\textsf{car}_{(#1)}(\scri)}
\newcommand{\Ccelg}[1]{\mathcal{C}^\textsf{cel}_{#1}(S,\g)}
\newcommand{\Ccarg}[1]{\mathcal{C}^\textsf{car}_{(#1)}(\scri,\g)}
\newcommand{\dCar}{\delta}

\renewcommand{\l}{\ell}

\newcommand{\ada}{\mathrm{ad}_A}

\newcommand{\T}[1]{T_{#1}}

\renewcommand{\a}[1]{\alpha_{#1}}
\newcommand{\Aa}[1]{\mathsf{a}_{#1}}
\newcommand{\ap}[1]{\alpha'_{#1}}
\newcommand{\Aap}[1]{\mathsf{a}'_{#1}}
\newcommand{\app}[1]{\alpha''_{#1}}

\newcommand{\Ac}[1]{A^{[#1]}}

\newcommand{\Pairg}[1]{\big\langle #1\big\rangle_\g}

\newcommand{\scri}{\mathrsfso{I}}

\newcommand{\Qa}{Q_\alpha}

\newcommand{\Qad}{\dot{Q}_\alpha}

\newcommand{\cq}{\mathscr{Q}}
\newcommand{\cqAas}{\mathscr{Q}_{\text{\scalebox{1.24}{$\Aa{s}$}}}}
\newcommand{\q}[2]{\hq_{#1}^{\scriptscriptstyle{\!(#2)}}}
\newcommand{\cQ}[2]{Q^u_{#1}\left[#2\right]}

\newcommand{\soft}[1]{#1^{\mathsf{S}}}
\newcommand{\hard}[1]{#1^{\mathsf{H}}}

\newcommand{\lbr}{\llbracket}
\newcommand{\rbr}{\rrbracket}
\newcommand{\poisson}[1]{\big\{#1\big\}}

\newcommand{\cyc}{\overset{\circlearrowleft}{=}}

\newcommand{\bfm}[1]{\boldsymbol{#1}}

\newcommand{\bfa}{\bfm{\alpha}}

\usepackage{tikz}

\definecolor{myorange}{RGB}{223, 109, 20}

\definecolor{argile}{RGB}{239, 239, 239}
\definecolor{beige}{RGB}{254, 253, 240}

\begin{document}

\title{\Large{\bf Asymptotic Higher Spin Symmetries III:\\
Noether Realization in Yang-Mills Theory}}

\author{Nicolas Cresto\,\orcidlink{0009-0006-7263-8777}$^{1,2}$\thanks{ncresto@perimeterinstitute.ca}}
\date{\small{\textit{
$^1$Perimeter Institute for Theoretical Physics,\\ 31 Caroline Street North, Waterloo, Ontario, N2L 2Y5, Canada\\ \smallskip
$^2$Department of Physics \& Astronomy, University of Waterloo,\\Waterloo, Ontario, N2L 3G1, Canada
}}}

\maketitle
\begin{abstract}
We construct a non-perturbative action of the higher spin symmetry algebra on the asymptotic Yang-Mills phase space. 
We introduce a symmetry algebroid which admits a realization on the asymptotic phase space generated by a Noether charge defined non-perturbatively for all spins. 
This Noether charge is naturally conserved in the absence of radiation. 
Furthermore, the algebroid can be restricted to the covariant wedge symmetry algebra, which we analyze for non radiative cuts. 
The key ingredient in this construction is to consider field and time dependent symmetry parameters constrained to evolve according to equations of motion dual to (a truncation of) the asymptotic Yang-Mills equations of motion.
This result then guarantees that the underlying symmetry algebra is
represented canonically as well.\\

\ni \textbf{Keywords:}
Yang-Mills, Non-Abelian Gauge Theory, Asymptotic Symmetries, Celestial Holography, Wedge Algebra, Covariant Wedge Algebra, Noether Representation, Twistor Theory.
\end{abstract}

\newpage
\tableofcontents
\newpage

\section{Introduction}

The study of asymptotic symmetries for abelian and non-abelian gauge theories is much more recent that its gravitational counterpart which dates from the 1960' \cite{Bondi:1962px, Sachs:1962wk}.
This illustrates one of these situations where the quest for Quantum Gravity led to new discoveries and perspectives in non-gravitational theories.
The starting point was the realization that the leading soft \textit{graviton} theorem \cite{PhysRev.140.B516} could be reinterpreted as a Ward identity for the supertranslation charge \cite{Strominger:2013jfa, He:2014laa}. 
This relation between soft theorems and asymptotic charges is \textit{generic} and does not rely on the peculiarities of the gravitational analysis.
On top of that, the position space version of the leading soft graviton theorem was proven to be equivalent to the leading displacement memory effect \cite{Strominger:2014pwa}.
We refer the reader to the lecture notes \cite{Strominger:2017zoo} and the introductions of our companion papers \cite{Cresto:2024fhd, Cresto:2024mne} for an ample list of references.
The main point is that an \textit{equivalence} between three seemingly disconnected topics related to infrared physics was unraveled for General Relativity (GR). 
Nowadays known as \textit{infrared} (IR) \textit{triangle}, it was then natural to look for a similar set of relationships in the case of gauge theories.

Historically, in 2014, \cite{He:2014cra} established the equivalence between leading soft \textit{photon} theorem (see \cite{PhysRev.96.1428, PhysRev.96.1433, low1958bremsstrahlung, kazes_generalized_1959, yennie_infrared_1961, burnett1968extension} for the original references on the leading and sub-leading soft photon theorems) and the Ward identity for super-phase rotation $\Aa0(z,\bz)$---the so-called large gauge transformation of \cite{He:2014cra}.\footnote{See also \cite{Kapec:2015ena} for the generalization to massive charged particles.} 
Later that year, \cite{Lysov:2014csa} studied the implications of the subleading soft photon theorem and how it could be formulated as the Ward identity for a new set of charges associated to a vector onto the celestial sphere (see \cite{Laddha:2017vfh} for the generalization to massless particles and \cite{Hirai:2018ijc} for the extension to massive scalar QED and a discussion of the associated memory effects).
In our analysis, we shall denote this vector by $\Aa1(z,\bz)$.
The generalization to non-abelian gauge theories was developed in parallel:
\cite{Strominger:2013lka, He:2015zea} proved that one can re-interpret the leading soft \textit{gluon} theorem as the Ward identity for the super-phase rotation (in color space) symmetry.
The associated color memory effect was introduced in \cite{Pate:2017vwa} and the IR $S$-matrix was then further constrained by the discovery of the subleading soft gluon theorem \cite{Casali:2014xpa, Broedel:2014fsa, SabioVera:2014mkb}.
Exactly as in the abelian case, the latter can be recast as the Ward identity for a charge parametrized by the vector $\Aa1$ (now $\g$-valued, with $\g$ the gauge Lie algebra) \cite{Adamo:2015fwa}.

All the preceding results hold for the tree-level $S$-matrix. 
See \cite{He:2014bga} for one-loop correction to the subleading soft theorems and \cite{Feige:2014wja} for factorization properties of scattering amplitudes to all orders. 
See also \cite{Sahoo:2018lxl, Campiglia:2019wxe, AtulBhatkar:2020hqz} for the loop and logarithmic corrections to the soft theorems.

Let us emphasize that an equivalent way to phrase the various Ward identities aforementioned is to say that the $S$-matrix is invariant under an infinitesimal transformation generated by a charge which contains a soft and a hard part. 
The soft part, linear in the gauge field, only exists for local (i.e. not constant) large gauge transformations. 
It is related to the insertion of the soft mode while the hard part, quadratic in the gauge field, acts on the in- and out-states to create the soft factors appearing in soft theorems \cite{Strominger:2017zoo}.

However, how to relate the charges associated to the sub-leading soft theorem in gauge theories (or the sub-sub-leading in GR) to asymptotic symmetries was not immediately transparent.
In 2016, Campiglia and Laddha \cite{Campiglia:2016hvg} proposed to interpret the sub-leading soft theorem as coming from the Ward identity associated to a charge parametrized by an over-leading, linear in $r$, gauge transformation---with $r$ the radial coordinate which tends to infinity on $\scri$.
In that context, the vector on the sphere previously considered by Strominger et al. then becomes the derivative onto the sphere of this over-leading gauge transformation.
Even if such a divergent gauge transformation does not preserve the asymptotic fall-offs of the gauge field and leads to divergent charges, there is still a natural finite part of the charge that one can relate to the sub-leading soft theorem.
The prescription is to neglect the divergent piece if the latter corresponds to the smearing of the \textit{leading} charge aspect with the divergent parameter.
This is an important point, also emphasized in \cite{He:2019pll}---where the analysis of \cite{Campiglia:2016hvg} is extended to the non-abelian case.
Indeed, this is because we know that the leading soft photon/gluon theorem holds and relates to the conservation of the leading charge that we can neglect the divergent contribution.
The fact that the divergent gauge transformation enters in a charge associated to a ``more leading'' soft theorem than the one associated to the finite part of the charge is key to get a finite Ward identity.
\medskip

The present work deals with asymptotic \textit{higher} spin symmetries, which in the context of gauge theories means that the symmetry parameters $\Aa{s}$ and the associated charges are labeled by their spin-weight $s\in\N$, hence generalizing the aforementioned $\Aa0$ and $\Aa1$.
In 2018, \cite{Hamada:2018vrw} unraveled a set of Ward identities constraining the sub${}^s$-leading infrared structure of gauge theory (and gravity).
This was understood \cite{Li:2018gnc} as the universal\footnote{On top of the loop and log corrections already mentioned, the sub${}^s$-leading soft theorems---$s\geq 2$ in gauge theory---are no longer universal and depend upon the details of the bulk interactions \cite{Li:2018gnc}.} tree-level part of sub${}^s$-leading soft theorems, thus giving information on the higher order soft behavior of gauge bosons and gravitons.
Thanks to the IR triangle, one knows that an interpretation of this result in terms of asymptotic symmetries should be possible.
In particular, \cite{Campiglia:2018dyi} conjectured\footnote{And showed for the sub-sub-leading case.} for massless QED how to relate these newly found sub${}^s$-leading soft theorems to the conservation of asymptotic charges associated to over-leading $\mathcal{O}(r^s)$ gauge parameters.
We expect the latter to be related to our parameters $\Aa{s}$ but this link is not the point of the present paper.

Importantly, asymptotic higher spin symmetries also appeared in the celestial holography context (see \cite{Pasterski:2021raf} and references therein for a review). 
There, the main objects of interest are celestial amplitudes/operators, obtained after a Mellin transform in energy of the usual momentum space amplitudes/operators.
The \textit{energetically} soft theorems described so far then morph into \textit{conformally} soft theorems, where the soft energy limit of the boson is replaced by a limit in the conformal weight of the associated conformal operator \cite{Pate:2019mfs}.
Equivalently, one can study conformally soft theorems and their Ward identities by considering 2-operators celestial operator product expansions (OPEs) where one operator is the soft current \cite{Fan:2019emx}.
The simple poles on the sphere appearing in celestial amplitudes and OPEs come from the collinear singularities in momentum space scattering amplitudes.
These collinear singularities can be studied on their own via Feynman diagrams.
However, one can bypass the diagrammatic approach and show that the 2-points celestial OPE coefficients are fully constrained by the asymptotic symmetries associated to the conformally leading and sub-leading soft gluon theorems (and similarly for GR including the sub-sub-leading order) \cite{Pate:2019lpp}.
In the same way \cite{Hamada:2018vrw, Li:2018gnc} discovered a whole tower of sub${}^s$-leading energetically soft theorems, \cite{Guevara:2021abz, Strominger:2021mtt} studied a tower of sub${}^s$-leading conformally soft currents, characterized by definite conformal weights, whose OPEs satisfy respectively the $\cS$-algebra, $w_{1+\infty}$ algebra and $\cS w_{1+\infty}$ algebra\footnote{Technically the wedge sub-algebra of the loop algebra of these.} for pure Yang-Mills (YM), pure GR and Einstein-Yang-Mills theory.
See also \cite{Himwich:2023njb} for an excellent account of the equivalence between momentum-space soft limits and emissions of particles of definite conformal weight; and how the $w_{1+\infty}$ soft algebra organizes the universal part of the infinite tower of soft theorems, both with massless and massive particles.\footnote{See \cite{Elvang:2016qvq, Melton:2022fsf} for the deformation of the soft gluon theorems and the associated soft algebra in the presence of a massless scalar field.} 

Finally, \cite{Freidel:2021ytz, Freidel:2023gue} identified a particular truncation in the asymptotic Einstein's equations and YM equations of motion in which the Weyl tensor, respectively field strength, was recast as a collection of higher spin fields $\tQ_s$.
The latter were shown to satisfy a simple set of evolution equations, which represents the input of the analysis of the current work, cf.\,\eqref{EOMYM}.
The conservation of these higher spin `charges' (i.e. the fact that they commute with the $S$-matrix) truncated to quadratic order was then equivalent to the infinite tower of conformally soft theorems.
In this approach, the label `charges' was justified as the (renormalized version of) $\tQ_s$ reproduced the celestial $w_{1+\infty}$ and $\cS$-algebras generated by the increasingly subleading soft modes of \cite{Guevara:2021abz, Strominger:2021mtt}, thus furnishing a phase space representation of this tower of symmetries.
Nevertheless, a derivation à la Noether for the entire tower of charges was missing, as well as a non-perturbative treatment.
The present paper remedies to this in the case of YM theory, while \cite{Cresto:2024fhd, Cresto:2024mne} dealt with General Relativity.

We also refer the reader to \cite{Campiglia:2021oqz, Nagy:2022xxs, Nagy:2024dme, Nagy:2024jua} for an analysis of over-leading gauge transformations in YM theory and gravity and the definition of an extending phase space compatible with such transformations.
\medskip

Hence the goal of this work is to extend the gravitational analysis of asymptotic higher spin symmetries of our companion papers \cite{Cresto:2024fhd, Cresto:2024mne} to non-abelian gauge theory.
We use a set of Carrollian symmetry parameters $\a{s}$ that satisfy certain evolution equations \eqref{dualEOMYM} in order to define non-perturbatively Noether charges that realize a \hyperref[theoremSAlgebroid]{$\cS$-algebroid} on the asymptotic phase space.
As a by-product of this general result, we construct for all spin $s$ the Noether charge associated to the celestial symmetry parameter $\Aa{s}$, where the latter now plays the role of initial condition for the time evolution of $\a{s}$.
We also discuss in details when does the algebroid reduce to an algebra, the so-called \textit{covariant wedge}, which defines the symmetry algebra of a non-radiative cut of $\scri$.
Besides, as we did as the end of \cite{Cresto:2024mne}, we present a complimentary standpoint to our main approach which makes the relation to twistor theory more patent.

The paper is organized as follows:
We start in section \ref{secYM:Preliminaries} with a reminder of the notation introduced in the first two parts of this work, together with the defining evolution equations for the YM higher spin charges aspects $\tQ_s$.
In section \ref{secYM:Master}, we define the master charge together with the dual equations of motion (EOM) that $\a{}$ has to satisfy and construct the symmetry transformation of the asymptotic gauge potential $A$.
We show that this transformation is the realization of the $\cS$-algebroid in section \ref{secYM:algebroid}.
We relate the master charge to the Noether charge in section \ref{secYM:Noether} and give an explicit solution of the dual EOM in section \ref{secYM:solEOM}.
This allows to relate Carrollian and celestial symmetry parameters.
Next, in section \ref{secYM:RenormCharge}, we show how to algorithmically and non-perturbatively build a renormalized charge aspect for each spin which is conserved in time in the absence of radiation.
We then dedicate section \ref{secYM:ActionOnA} to the computation of the hard action on $A$.
We introduce in section \ref{secYM:wedge} the covariant wedge algebra, both on $\scri$ and on the sphere.
Section \ref{secYM:twistor} shows how to relate the graded vector $\a{}$ to a function $\fa$ which naturally appears in twistor space.
A collection of appendices complements the main text with details of demonstrations.

\section{Preliminaries \label{secYM:Preliminaries}}

We follow the conventions introduced in \cite{Cresto:2024fhd, Cresto:2024mne} and work on $\scri=\R\times S$, with $S$ a 2-dimensional complex manifold with a
complex structure. 
$(u,r,z,\bz)$ denote the Bondi coordinates and we use the null
dyad $(m,\bm)$ on the sphere $S$, which can be a regular or a punctured sphere.\footnote{So that we often refer to $S$ as the `sphere' even though we do not fix the topology.}
We denote $D=m^A D_A$ the covariant derivative along $m$ preserving the complex structure.
The indices $(A,B,\ldots)$ live on $S$ and thus run over 1 and 2.
$\bfm{\epsilon}_S$ is the area form on $S$.

\subsection{Symplectic potential and equations of motion}

We study non-abelian gauge theory over flat space, with the associated Lie algebra $\g$, and denote by $A_\mu$ and $F_{\mu\nu}$ the gauge potential and field strength respectively.
As we did in our GR analysis, we work with spin-weighted scalars rather than tensors over the sphere.
As an example, when performing a $1/r$ expansion around null infinity of the YM gauge field \cite{Strominger:2017zoo}, where for instance $A_z$ takes the form
\begin{equation}
    A_z=\sum_{n=0}^\infty\frac{A_z^{(n)}(u,z,\bz)}{r^n},
\end{equation}
then we can show that only $A_z^{(0)}$ is part of the asymptotic phase space and we thus denote its contraction with the frame field by $A\equiv A^{(0)}_B m^B$.
Similarly, the canonically conjugated variables will be denoted by $\news=\pa_u A$ and its complex conjugate $\bnews$.
Notice also that we work with Lie algebra valued quantities to lighten the notation, i.e. $A_\mu=A_\mu^b Z_b$ with $\{Z_b\}$ a basis of anti-hermitian Lie algebra generators. 
Besides, we adopt the convention that the complex conjugation acts only on the spacetime components, i.e. relates to the frame fields, and not on the algebra generators (in case the latter are complex).
Even if we shall write expressions such as $\bA$, the latter must then be understood as $\bA=\bA^bZ_b=A_B^{(0)b}\bm^B Z_b$.

$A$ and $\news$ parallel the role of the shear and news in asymptotically flat spacetimes.
This is transparent since the asymptotic YM symplectic potential is given by\footnote{$\Pairg{Z_a,Z_b}=-\Tr(Z_aZ_b)$, where $\Tr$ denotes the trace, i.e. the Cartan–Killing form for $\g$, normalized such that $\Tr(Z_aZ_b)=-\delta_{ab}$ (recall that we use anti-hermitian generators). \label{footPairg}}
\begin{equation}
\Theta=\frac{2}{\gYM^2}\int_{\scri}\rd u\wedge\bfm{\epsilon}_S \,\Pairg{\bnews,\delta A}, \label{SymPotYM}
\end{equation}
which is the holomorphic Ashtekar-Streubel asymptotic symplectic potential\footnote{The Ashtekar-Streubel symplectic potential is real and takes the form 
\begin{equation}
\Theta^{\mathsf{AS}}=\frac1{\gYM^2} \int_{\scri}\Big(\Pairg{\bnews,\delta A}+\Pairg{F,\delta\bA}\Big).
\end{equation}
If we add a total variation together with a boundary term, namely $\frac{1}{\gYM^2}\big(\delta\left(\int \langle\bnews, A\rangle_\g\right)- \int \pa_u\langle A,\delta\bA\rangle_\g\big)$, we transform it into its holomorphic version.} \cite{ashtekar1981symplectic}.

By performing an asymptotic expansion around $\scri$ and relating the field strength to the higher spin charges aspects $\tQ_s$, $s\geqslant -1$, the YM EOM in vacuum are approximated by \cite{Freidel:2023gue}
\begin{equation}
    \boxed{\pa_u\tQ_s=D\tQ_{s-1}+\big[A,\tQ_{s-1}\big]_\g},\qquad s\geqslant 0. \label{EOMYM}
\end{equation}
We define $\Dcal$ the asymptotic gauge covariant derivative on the sphere,
\begin{equation}
    \boxed{\Dcal:=D+\ada}, \label{defDcalYM}
\end{equation}
where $\ada(\cdot)=[A,\cdot]_\g$ is the adjoint action in the Lie algebra $\g$.
The EOM are thus expressed as $\pa_u\tQ_s=\big(\Dcal\tQ\big)_s\equiv\Dcal\tQ_{s-1}$.
In a similar fashion as for GR, the initial condition $\tQ_{-1}=-\bnews$ encodes the radiative content of the gauge potential.
We let for future work the detailed account of the YM EOM written in terms of the higher spin charges.
Here we take \eqref{EOMYM} as a defining property for $\tQ_s$ and it is sufficient to know that the evolution equations are the exact asymptotic EOM for $s=0,1$ and then become an approximation.

\subsection{Graded vector spaces and symmetry parameters}

As we did in our companion papers \cite{Cresto:2024fhd, Cresto:2024mne}, it is convenient to classify fields by their Carrollian or celestial weights and form associated graded vector spaces.

As a reminder, a Carrollian field $\Phi(u,z,\bz) \in \Ccar{\dCar,s}$ of Carrollian weight $\delta$ and spin-weight (or helicity) $s$ transforms under supertranslations as $\delta_T\Phi= T\pa_u \Phi$ and under sphere diffeomorphisms as\footnote{$\mathcal{Y}=\mathcal{Y}^B\pa_B=YD+\bY\bD$.}
\be
\delta_{\mathcal Y}  \Phi 
= \big(YD + \tfrac12 (\delta  + u\pa_u + s) DY   \big)\Phi
+ \left(\bY\bD+\tfrac12 (\delta + u\pa_u  - s)\bD\bY \right) \Phi .
\label{sDeltaDefCarroll}
\ee
This transformation amounts to the usual definition of a conformal primary field in 2d \cite{francesco2012conformal, Barnich:2021dta, Donnay:2021wrk} if $Y$ ($\bY$) is (anti-)holomorphic.
See \cite{Raclariu:2021zjz} for their relevance in celestial holography.
Here we consider the extension to arbitrary sphere diffeomorphisms \cite{Freidel:2021dfs}. 
In the present paper we work in flat space, so that supertranslations reduce to global time translation and sphere diffeomorphisms reduce to M\"obius transformations, i.e. global conformal transformations of the sphere.
In the following we will leverage the natural projection between Carrollian and celestial fields given by the evaluation at the cut $u=0$:\footnote{Namely a choice of embedding of $S$ into $\scri$.}
\be 
\Ccar{\dCar,s} & \to \Ccel{(\dCar,s)} \cr 
\Phi &\mapsto \phi(z,\bz) = 
\Phi(u=0,z,\bz).
\ee 

Recall that $D$ and $\pa_u$ are operators of weights $(1,1)$ and $(1,0)$ respectively.
From \eqref{defDcalYM}, it is thus clear that $A\in\Ccarg{1,1}$.
Moreover, since the initial charge $\tQ_{-1}=-\bnews\in\Ccarg{2,-1}$, we deduce that $\tQ_s\in\Ccarg{2,s}$.
Using the insight of our GR analysis, we introduce $\g$-valued time dependent symmetry parameters $\a{s}\in\Ccarg{0,-s}$ and we denote the series $(\a{s})_{s\in\N}$ by $\a{}$.
In other words, $\a{}$ is a graded vector such that
\begin{equation}
    \a{}\in\tV(\scri,\g), \qquad \tV(\scri,\g):=\bigoplus_{s=0}^\infty\tV_s\quad\textrm{with}\quad \tV_s\equiv\Ccarg{0,-s}.
\end{equation}
The filtration associated to the gradation of $\tV(\scri,\g)$ is denoted by
\begin{equation}
    \tV^s\equiv\tV^s(\scri,\g):= \bigoplus_{n=0}^s\tV_n.
\end{equation}
We also introduce the celestial counterpart of $\a{s}$, denoted $\Aa{s}$, where
\begin{equation}
    \Aa{}\in\V(S,\g), \qquad \V(S,\g):=\bigoplus_{s=0}^\infty\V_s\quad\textrm{with}\quad \V_s\equiv\Ccelg{(0,-s)}.
\end{equation}

\section{Master charge and symmetry transformation \label{secYM:Master}}

We define the smeared charges\footnote{$\int_S\equiv\int_S\bfm{\epsilon}_S$ and the product $\Tr\big(\tQ_s\a{s}\big)\in\Ccar{2,0}$ is a scalar density.}
\begin{equation}
\cQ{s}{\a{s}}:=\frac{2}{\gYM^2}\int_S\Pairg{\tQ_s,\a{s}}(u,z,\bz). \label{Qsas}
\end{equation}
Since $\cQ{s}{\a{s}}$ involves an integral over the sphere (at $u=\mathrm{cst}$), so that we can freely integrate by parts.
Using \eqref{defDcalYM} to construct 
\begin{equation}
    (\Dcal\a{})_s=\Dcal\a{s+1}:=(D+\ada)\a{s+1},
\end{equation}
we get that\footnote{$\Tr\big(\tQ_s[A,\a{s+1} ]_\g\big)=-\Tr\big([A,\tQ_s]_\g\a{s+1}\big)$ and we assume that $\sum_s\tQ_s\a{s}$ is regular at the punctures if $S$ is not the regular sphere.}
\begin{equation}
    \cQ{s}{\Dcal\a{s+1}}=-\Dcal\cQ{s}{\a{s+1}}, \label{QDDQ}
\end{equation}
where
\begin{equation}
    \Dcal\cQ{s}{\a{s+1}}=\frac{2}{\gYM^2}\int_S\Pairg{(\Dcal\tQ)_{s+1},\a{s+1}}.
\end{equation}

We then construct the \textit{master charge} $\Qa^u$,
\begin{equation}
\boxed{\Qa^u := \sum_{s=0}^{\infty} \cQ{s}{\a{s}}}. \label{masterChargeYM}
\end{equation}
Using \eqref{EOMYM} and \eqref{QDDQ}, its time evolution takes the form
\begin{equation}
\pa_u\Qa^u = \sum_{s=0}^\infty \Big(\cQ{s}{\pa_u \a{s}} - \cQ{s-1}{\Dcal\a{s}}\Big) = -\cQ{-1}{\Dcal \a0}+\sum_{s=0}^\infty Q^u_{s}\big[\pa_u \a{s} - \Dcal\a{s+1}\big].
\end{equation}
By imposing what we shall refer to as the dual EOM, 
\begin{equation}
\boxed{\pa_u\a{s}\equiv\dot\alpha_s= \Dcal\a{s+1}}, \quad s\geqslant 0, \label{dualEOMYM}
\end{equation}
and remembering that $\tQ_{-1}=-\bnews$, we obtain that 
\begin{equation}
\Qad^u = \bnews\big[\Dcal\a0\big]=-\bnews\big[\da A\big], \label{Qadot}
\end{equation}
where we have defined the transformation\footnote{This choice of sign implies that the anchor $\delta$ \cite{algebroid} is an anti-homomorphism of Lie algebroids, like the one we defined for GR in \cite{Cresto:2024mne}. 
Sign consistency is especially important when coupling GR and YM.}
\begin{equation}
\boxed{\da A:= -\Dcal\a0}.  \label{deltaA}
\end{equation}

The master charge is conserved in the absence of left-handed gluons, namely $\Qad^u=0$ when $\bnews=0$.
Moreover, by imposing the fall-off condition $\Qa^{+\infty}=0$,\footnote{Namely $\lim_{u\to\infty}\tQ_s\a{s}=0$.} it takes the form of an integral over a portion of $\scri$,
\begin{equation}
    \Qa^u=\int_u^{\infty}\bnews\big[\da A\big]. \label{Qau}
\end{equation}
Notice that $\Qa^{-\infty}\propto\int_{\scri}\Pairg{\bnews,\da A}$. 
In other words $\Qa^{-\infty}$ can be constructed from the symplectic potential \eqref{SymPotYM}.
Hence, the dual EOM are key in order for $\Qa^{-\infty}$ to be a Noether charge, see Sec.\,\ref{secYM:Noether}.

\section{$\cS$-algebroid \label{secYM:algebroid}}

In this section, we prove that the $\alpha$-transformation \eqref{deltaA} is a realization on the asymptotic Yang-Mills phase space $\PS$ of an algebroid extension of the $\cS$-algebra \cite{Guevara:2021abz}. 

Let us first show that one can build a bracket $\lbr\alpha,\alpha'\rbr$ such that\footnote{$\big[\da,\dap\big]$ is the Lie bracket of vector fields over fields space \cite{Freidel:2020xyx, saunders1989geometry} and \cite{Cresto:2024mne}.}
\begin{equation}
\boxed{\big[\da,\dap\big]A=-\delta_{\lbr\alpha,\alpha'\rbr}A}. \label{SalgebraMorphism}
\end{equation}
From the transformation \eqref{deltaA}, we deduce that
\begin{align}
\da\dap A &=-\da\big(D\ap0+[A,\ap0]_\g\big) \nn\\
&=-D\da\ap0-\big[A,\da\ap0\big]_\g+\big[\Dcal\a0,\ap0\big]_\g \\
&=-\Dcal(\da\ap0)+\big[\Dcal\a0,\ap0\big]_\g. \nn
\end{align}
The action of the commutator $\big[\da,\dap\big]$ on $A$ is thus
\begin{equation}
    \big[\da,\dap\big]A =-\Dcal\big(\da\ap0-\dap\a0\big) +\big[\Dcal\a0,\ap0\big]_\g- \big[\Dcal\ap0,\a0\big]_\g.
\end{equation}
Merely using the Leibniz rule on the third term, we get
\begin{equation}
    \big[\da,\dap\big]A=\Dcal \big([\a0, \ap0]_\g+\dap\a0-\da\ap0\big)= \Dcal\lbr\alpha,\alpha'\rbr_0,
\end{equation}
for 
\begin{equation}
    \lbr\alpha,\alpha'\rbr_0=[\a0,\ap0]_\g+\dap\a0-\da\ap0. \label{Sbracket0}
\end{equation}
Notice that the Leibniz rule for the derivative operator $\Dcal$ on the Lie algebra bracket $[\cdot\,,\cdot]_\g$ is nothing else than the Jacobi identity for this bracket.
Indeed,
\begin{equation}
    \Dcal[\a{s},\ap{n}]_\g=[D\a{s},\ap{n}]_\g+[\a{s},D\ap{n}]_\g+\big[A,[\a{s},\ap{n}]_\g\big]_\g=[\Dcal\a{s},\ap{n}]_\g+[\a{s},\Dcal\ap{n}]_\g, \label{DcalLeibniz}
\end{equation}
after using Jacobi.
Since the evolution in time of $\a0$ is related to $\a1$ via \eqref{dualEOMYM}, we have to define a bracket of degree 1 which respects this property as well.
By induction, we thus need to construct $\lbr\a{},\ap{}\rbr_s$, for all $s\in\N$.

Hence we generalize the bracket \eqref{Sbracket0} for $s\geqslant 0$ by writing
\begin{equation} \label{Sbracket}
    \boxed{\lbr\alpha,\alpha'\rbr= [\alpha,\alpha']+\dap\alpha-\da\alpha'},
\end{equation}
where the evaluation at degree $s$ satisfies
\begin{equation}
    [\alpha,\alpha']_s=\sum_{n=0}^s\big[\a{n} ,\ap{s-n}\big]_\g\qquad\textrm{and}\qquad (\da\alpha')_s=\da\alpha'_s.
\end{equation}
Notice that $[\alpha,\alpha']$ is the natural extension of the Lie algebra bracket $[\cdot\,,\cdot]_\g$ to the full graded vector space $\tV(\scri,\g)$.
Indeed, using the inclusion map $\iota:\tV_s\to\tV(\scri,\g)$, such that $\a{s}\mapsto (0,\ldots,0,\a{s},0,\ldots)$, we see that
\begin{equation}
    \big[\iota(\a{s}),\iota(\ap{s'})\big]_n= \left\{\begin{array}{ll}
        [\a{s},\ap{s'}]_\g & \textrm{if }n=s+s', \\
        0 & \textrm{otherwise}.
    \end{array}\right.
\end{equation}
This also means that $[\cdot\,,\cdot]$ is a bracket of degree 0.

Since the Carrollian symmetry parameters $\alpha$ are restricted to follow the dual EOM \eqref{dualEOMYM}, we define the $\sfS$-space accordingly.

\begin{tcolorbox}[colback=beige, colframe=argile]
\textbf{Definition [$\sfS$-space]}
\be 
\sfS :=\Big\{\,\alpha\in\tV(\scri,\g)~\big|~\pa_u\a{s}=\Dcal\a{s+1},~  s\geqslant 0~ \Big\}.
\ee 
\end{tcolorbox}

\ni The fact that the parameters $\a{s}$ are constrained to follow an evolution equation involving $A$ implies that they are field dependent.
It is thus natural and necessary that the $\sfS$-bracket \eqref{Sbracket} involves the variations $\da\ap{}$ and $\dap\a{}$.
This is a well-known fact in the algebroid literature \cite{algebroid}.
We refer the reader to \cite{Cresto:2024mne} for a concise reminder of the key elements about Lie algebroids (see also \cite{algebroid, Barnich:2010xq, Ciambelli:2021ujl} for a more general discussion, as well as \cite{Barnich:2010eb}).
The following theorem then represents one of the main result of this paper.

\begin{tcolorbox}[colback=beige, colframe=argile] \label{theoremSAlgebroid}
\textbf{Theorem [$\cS$-algebroid]}\\
The space $\cS\equiv\big(\sfS,\lbr\cdot\,,\cdot\rbr,\delta\big)$ equipped with the $\sfS$-bracket \eqref{Sbracket} and the anchor map $\delta$,
\begin{align}
    \delta : \,\,&\cS\to \X(\PS) \nn\\
     &\alpha\mapsto\da, \label{defdeltaYM}
\end{align}
is a Lie algebroid over $\PS$.
This means that the anchor map is a morphism of Lie algebroid, namely $\big[\da,\dap\big]A=-\delta_{\lbr\alpha,\alpha'\rbr}A$, cf. \eqref{SalgebraMorphism} and also that the $\sfS$-bracket closes, i.e. $\pa_u\lbr\a{},\ap{}\rbr= \Dcal\lbr\a{},\ap{}\rbr$.
\end{tcolorbox}

\paragraph{Proof:}
First of all, the proof that the $\sfS$-bracket satisfies the Jacobi identity follows from a general result about symmetry algebroids, see \cite{Cresto:2024mne,algebroid}.
In words, since $[\cdot\,,\cdot]$ is built from the Lie bracket $[\cdot\,,\cdot]_\g$---and thus satisfies Jacobi on its own---and that the anchor $\da$ acts as a derivative operator on $[\cdot\,,\cdot]$, i.e. $\da[\cdot\,,\cdot]=[\da\cdot\,,\cdot]+[\cdot\,,\da\cdot]$, then the Jacobi identity for $\lbr\cdot\,,\cdot\rbr$ is guaranteed to hold.
For the sake of completeness and to illustrate this general fact, we give the explicit demonstration in App.\,\ref{App:JacobiYM} (this result even holds over the whole space $\tV(\scri,\g)$).

Second of all, we already proved that $\delta$ is a Lie algebroid anti-homomorphism, cf. \eqref{SalgebraMorphism}, and thus an anchor map.

Third of all, in App.\,\ref{App:dualEOMYM}, we report the technical details to show that $\lbr\cdot\,,\cdot\rbr:\sfS\times\sfS\to\sfS$ is a well-defined bracket over $\sfS$, i.e. that it satisfies the dual EOM,
\begin{equation}
    \boxed{\pa_u\lbr\a{},\ap{}\rbr= \Dcal\lbr\a{},\ap{}\rbr}. \label{dualEOMYMbracket}
\end{equation}
The essence of the proof goes as follows:
We mentioned already that $\Dcal$ is the asymptotic gauge covariant derivative onto $S$.
Its appearance in the current analysis is thus natural.
However, $\Dcal$ may or may not be a \textit{derivative operator} depending on which bracket is acts on.
Hence, we compute the potential Leibniz rule anomaly of $\Dcal$ (and similarly for $\pa_u$) on the various brackets.
This then allows us to straightforwardly ensure that the bracket $\lbr\cdot\,,\cdot\rbr$ closes in $\sfS$.

Concretely, we define the Leibniz rule anomaly for any bracket $[\cdot\,,\cdot]$ and differential operator $\mathscr{D}$ as
\begin{equation}
    \cA\big([\cdot\,,\cdot],\mathscr{D} \big)=\mathscr{D}[\cdot\,,\cdot]-[\mathscr{D}\cdot\,,\cdot]-[\cdot\,,\mathscr{D}\cdot].
\end{equation}
We find that
\begin{equation} \label{DcalLeibniz1}
    \cA_s\big([\a{},\ap{}],\Dcal\big)= \big[\dap A,\a{s+1}\big]_\g-\big[\da A,\ap{s+1}\big]_\g
\end{equation}
and
\begin{equation}\label{DcalLeibniz2}
    \cA\big(\lbr\a{},\ap{}\rbr, \Dcal\big)=\delta_{\Dcal \a{}}\ap{}-\delta_{\Dcal \ap{}}\a{}.
\end{equation}
We can also compute
\begin{equation}\label{DcalLeibniz3}
    \cA\big(\lbr\a{},\ap{}\rbr, \pa_u\big)=\delta_{\pa_u\a{}}\ap{}-\delta_{\pa_u\ap{}}\a{}.
\end{equation}
Combining \eqref{DcalLeibniz2} with \eqref{DcalLeibniz3}, we obtain
\begin{equation}
    \cA\big(\lbr\a{},\ap{}\rbr, \pa_u-\Dcal\big)=\delta_{(\pa_u-\Dcal)\a{}}\ap{}-\delta_{(\pa_u-\Dcal)\ap{}}\a{}.
\end{equation}
If $\a{},\ap{}\in\sfS$, then the RHS of this last equation vanishes while its LHS reduces to $(\pa_u-\Dcal)\lbr\a{},\ap{}\rbr$.
Therefore
\begin{equation}
    \pa_u\lbr\a{},\ap{}\rbr=\Dcal \lbr\a{},\ap{}\rbr,
\end{equation}
so that $\lbr\a{},\ap{}\rbr\in\sfS$ if $\a{},\ap{}\in\sfS$, which concludes the proof of \eqref{dualEOMYMbracket} and of the theorem.

\section{Noether charge \label{secYM:Noether}}

In this section, we show that there exists a canonical representation of the higher
spin $\cS$-algebroid on the asymptotic YM phase space \eqref{SymPotYM}.
\textit{Mutatis mutandis}, this result follows exactly the same reasoning as the demonstration we provided in \cite{Cresto:2024mne}.
Consequently, we only emphasize the main arguments and refer the reader to our companion paper for details.

We choose the fields $(A,\bA)$ to belong to the Schwartz space \cite{schwartz2008mathematics}.
This hypothesis about the fall-off in $u$ is important for two reasons.
On one hand, at this stage of the analysis, we discard any boundary terms that arise when integrating by parts $\pa_u$ derivatives.
Therefore, the fields must tend to 0 when $|u|\to\infty$.\footnote{We let for future work a potential relaxation of this condition.
See \cite{Nagy:2024dme, Nagy:2024jua} for some recent developments.}
On the other hand, in section \ref{secYM:solEOM}, we construct solutions of the dual EOM \eqref{dualEOMYM} which are polynomials in $u$.
Hence, in order for $\tQ_s\a{s}$ to tend to 0 when $|u|\to\infty$, the gauge potential must decay faster than any positive inverse power of $u$, namely be part of the Schwartz space.

As in \cite{Cresto:2024mne}, the whole point of the demonstration is to show that one can build $\da\bnews$ such that 
\begin{equation}
    L_{\da}\Theta=0,
\end{equation}
where $L_{\da}$ is the Lie derivative in fields space along the vector field $\da$. 
Using Cartan calculus \cite{Freidel:2020xyx}, this amount to $I_{\da}\delta\Theta+\delta I_{\da}\Theta=0$,\footnote{$\delta$ here is the fields space exterior derivative, not to be confused with the anchor. 
By definition, both are related according to $I_{\da}(\delta A)=\da A$, for $I$ the fields space interior product \cite{Donnelly:2016auv}.} which means that
\begin{equation} \label{NoetherChargeYM}
    \boxed{I_{\da}\Omega=-\delta\Qa}\qquad \textrm{for}\qquad\boxed{\Qa:=\Qa^{-\infty}=I_{\da}\Theta}, 
\end{equation}
with $\Omega=\delta\Theta$ the symplectic form.
From \eqref{Qau}, the charge can be written as an integral over $\scri$,
\begin{equation}
    \boxed{\,\Qa=-\frac{2}{\gYM^2}\int_{\scri}\Pairg{\bnews,\Dcal\a0}\,}. \label{QaNoetherYM}
\end{equation}
The dependence on the higher spin $\a{s}$ is implicit via the dual EOM \eqref{dualEOMYM}.
The equation \eqref{NoetherChargeYM} shows that $\Qa$ is the \textit{Noether charge} for the $\cS$-algebroid action.
Using the morphism \eqref{SalgebraMorphism} and by definition of the Poisson bracket,\footnote{According to the symplectic form $\Omega$ (or the symplectic potential \eqref{SymPotYM}), we have that $\poisson{\bnews(x),A(x')}=\frac{\gYM^2}{2}\delta^3(x-x')$, where $x=(u,z,\bz)$.} we get that
\begin{equation}
    \boxed{\poisson{\Qa,O}=\da O} \qquad \textrm{and}\qquad \boxed{\poisson{\Qa, Q_{\ap{}}}=-Q_{\lbr\a{},\ap{}\rbr}},
\end{equation}
where $O$ is an arbitrary functional of $(A,\bA)$.

\section{Solution of the dual EOM \label{secYM:solEOM}}

In this section, we study the polynomial class of solutions to the dual EOM and give explicit expressions to all order in the $\gYM$ expansion.
We especially introduce the Lie algebroid map $\bfa$ which defines a unique $\bfa(\Aa{})\in\cS$ given an element $\Aa{}\in\V(S,\g)$.
This is essential to define a \textit{corner} symmetry charge with an associated \textit{celestial} symmetry parameter $\Aa{}$, namely a parameter living on the sphere.
Indeed, for now $\Qa$ is either defined over $\scri$ \eqref{QaNoetherYM}, using the $u$-dependent Carrollian symmetry parameter $\a0$, or as a corner integral \eqref{masterChargeYM} by summing over multiple $\a{s}$. 
The solution $\bfa(\Aa{})$ will allow us to write the Noether charge as a corner integral for a \textit{single} $\Aa{s}$ with definite spin-weight.

In order to define such a spin-$s$ Noether charge $\cqAas$, we need to understand the space of solutions of \eqref{dualEOMYM}.
Therefore, we now describe the filtration and the associated gradation of the $\cS$-algebroid.
The discussion follows the same steps as the GR analysis of \cite{Cresto:2024mne}.

\subsection{Filtration and gradation \label{secYM:filtration}}

$\cS$ has a natural filtration,
\begin{equation}\label{defFiltrationYM}
    \{0\} \subset\cS^0\subset \cS^1\subset \ldots\subset\cS^s\subset \ldots\subset \cS \quad\textrm{such that}\quad \cS=\bigcup_{n\in\N}\cS^n,
\end{equation}
where the $\cS^s=\cS\cap\tV^s$ are the subspaces for which $\a{n}=0$ for $n>s$.
This filtration is compatible with the bracket since
 \begin{equation}
    \lbr\cS^s,\cS^{s'}\rbr\subseteq \cS^{s+s'},\qquad s,s'\geqslant 0. \label{filtrationCompatibleYM}
 \end{equation}
The associated graded algebroid $\G{\cS}$ is defined as
\begin{equation}
\G{\cS}=\bigoplus_{s=0}^\infty\sfS_s,
\end{equation}
with 
\begin{equation}
    \sfS_0=\cS^0\qquad \textrm{and}\qquad \sfS_s=\cS^s/\cS^{s-1}\quad\textrm{for }s>0.  \label{defgradationYM}
\end{equation}
Each $\sfS_s$ is an equivalence class and we can write $\cS^s= \bigoplus_{n=0}^s\sfS_n$.
We can easily show that the following isomorphism holds:
\vspace{-0.2cm}
\be 
\sfS_s \simeq  \V_s. \label{IsoYM}
\ee 
To prove this, we just have to pick a representative element of $\sfS_s$ and show that it is fully determined by an element $\Aa{s}\in\V_s$.
The mapping goes as follows: 
For $\a{}\in\cS^s$, $\pa_u\a{s}=0$ so that $\a{s}$ is equal to its value at any cut of $\scri$.
For definiteness, we take the $u=0$ cut.
Therefore,
\be 
\a{s}=\Aa{s},\quad  \mathrm{where}\quad  \Aa{s}:= \a{s}|_{u=0}\in \V_s.
\ee
The other values, $\a{s-n}\neq 0$ for $1\leq n\leq s$ are determined recursively from $\a{s}$ by the dual equations of motion. 
We thus have constructed the unique section $\a{}$
\begin{align}
    \a{} :\,&\V_s\to \cS^s \nn\\
     &\, \Aa{s}\mapsto \a{}(\Aa{s}), \label{mapalpha}
\end{align} 
which is a solution of the dual EOM with initial condition given by
\be 
\a{s}(\Aa{s})\big|_{u=0}=\Aa{s}, \qquad \a{s-n}(\Aa{s})\big|_{u=0}=0,\quad \mathrm{for} \quad 1 \leq n \leq s.
\ee 
By construction, such $\a{}$ is precisely a representative element of $\cS^s/\cS^{s-1}$ and is fully given in terms of $\Aa{s}\in\V_s$, which proves \eqref{IsoYM}.
As we mentioned already, these solutions are polynomial in $u$ and thus diverge at infinity, which forces us to take the charge aspects $\tQ_s$ as part of the Schwartz space.

Since $\a{}:\V_s\to \cS^s$ is a linear map, we can extend it by linearity to the map $\bfa: \V(S) \to \cS$ on the full space,\footnote{We use the notation $\bfa(\Aa{})\equiv\a{}(\Aa{})$ interchangeably.}
\begin{align} \label{mapalphabis}
    \bfa :\,& \,
    \V(S,\g) \to \cS \nn\\
     & \,\, \Aa{} \mapsto \bfa(\Aa{}) := \sum_{s=0}^{\infty} \a{}(\Aa{s}).
\end{align}

\subsection{Explicit construction of the section $\alpha$}

We recast the dual evolution equation \eqref{dualEOMYM} with initial condition $\bfa(\Aa{})\big|_{u=0}=\Aa{}$ as follows:
\begin{equation}
    \boxed{\bfa{}(\Aa{})=\pui\Dcal\bfa{}(\Aa{}) +\Aa{}}. \label{solEOMYMcompact}
\end{equation}
More explicitly,
\begin{equation}
    \a{s}(\Aa{})=\big(\pui D+\pui[1]
    [A,\cdot]_\g\big)\a{s+1}(\Aa{})+\Aa{s},
\end{equation}
where it is understood that $\pui=\int_0^u$ acts on all products on its right.\footnote{The definition of $\pui$ is ambiguous and depends on a base point $\beta$ so that  $\pui O =\int_\beta^u \rd u' O(u')\rd u'$ and in general 
\begin{equation}
    \big(\pui[n]O\big)(u)=\int_\beta^u \rd u_1\int_\beta^{u_1}\rd u_2\ldots\int_\beta^{u_{n-1}}\rd u_n O(u_n).
\end{equation}
Here we choose $\beta=0$, namely the cut of $\scri$ where $\Aa{}$ is defined. 
See \cite{Cresto:2024mne} for a more detailed discussion.}
Therefore, the solution for $\a{s-1}$, $\a{s-2},\ldots$ and so on, is built by taking one more power of the operator $\pui\Dcal$ at each step.
Concretely, picking $\a{}\in\cS^s$, we can write
\begin{equation}
    \a{n}(\Aa{})=\sum_{k=0}^{s-n}\big(\pui \Dcal\big)^k\Aa{n+k}.
\end{equation}
To find a representative element of $\sfS_s$, we take only $\Aa{s}\neq 0$, cf. \eqref{IsoYM},
\begin{equation}
    \boxed{\a{n}(\Aa{s})=\big(\pui \Dcal\big)^{s-n}\Aa{s}}, \qquad 0\leq n\leq s.\label{solAlphanAs}
\end{equation}
This is a non-perturbative solution.
To illustrate this, let us see how it expands in power of $A$.
Using the formula
\begin{equation}
    (B+A)^k=\sum_{\l=0}^k\sum_{P=k-\l}B^{p_0}AB^{p_1}AB^{p_2}\ldots AB^{p_\l},\qquad\textrm{with}\quad P=\sum_{i=0}^\l p_i,
\end{equation}
which holds for arbitrary operators $A$ and $B$, we get that
\begin{align}
    \big(\pui\Dcal\big)^k=\left(\pui D+\pui \ada\right)^k &=\sum_{\l=0}^k\sum_{P=k-\l}\pui[p_0]D^{p_0}\bigg(\pui\Big[A,\pui[p_1]D^{p_1}\Big(\pui\Big[A,\pui[p_2]D^{p_2}\Big(\ldots \nn\\
    &\hspace{-0.6cm}\ldots \pui\Big[A,\pui[p_{\l-1}]D^{p_{\l-1}}\Big(\pui \Big[A,\pui[p_\l]D^{p_\l}\Big]_\g\Big)\Big]_\g\ldots \Big)\Big]_\g\Big)\Big]_\g\bigg). \label{puiDcalkYM}
\end{align}
Notice that each term in the sum over $\l$ contains $\l$ gauge fields $A$.
We shall use this way \eqref{solAlphanAs} and \eqref{puiDcalkYM} of writing the solution of \eqref{dualEOMYM} in the discussion about the renormalized charges in section \ref{secYM:RenormCharge}.
As an example, we give the explicit expressions of \eqref{solAlphanAs} when $\a{}\in\cS^2$, which implies that $\Aa{}=(\Aa0,\Aa1,\Aa2,0,\ldots)$:
\bs 
\begin{align}
    \a2(\Aa{}) &=\Aa2, \\
    \a1(\Aa{}) &=uD\Aa2+\Aa1+\pui[1][A,\Aa2]_\g, \\
    \a0(\Aa{}) &=\frac{u^2}{2}D^2\Aa2+uD\Aa1+ \Aa0+\pui[2]D[A,\Aa2]_\g \\*
    &+\pui\big[A,uD\Aa2+ \Aa1\big]_\g+\pui\big[A,\pui[1][A,\Aa2]_\g\big]_\g. \nn
\end{align}
\es

Let us emphasize that the construction of $\bfa(\Aa{})$ is algorithmic. 
In particular, the symmetry transformation $\da A=-\Dcal\a0$ is simply expressed in terms of $\a0$ but the latter incorporates a great deal of non-linearity and non-locality via the solution $\a0(\Aa{s})$.
From a perturbative point of view, the number of gauge fields appearing in the transformation $\d{s}{\Aa{s}}A\equiv\delta_{\bfa(\Aa{s})}A$ grows linearly with the helicity $s$.
However, from the perspective of the variable $\a{}$, the infinitesimal transformation is still $-\Dcal\a0$.
This is the key reason why our approach is \textit{non-perturbative}.
Moreover, since the proof that the action $\da A$ is canonically represented on the phase space holds for generic $\a{}$, we get for free that the \textit{non-linear} action $\d{s}{\Aa{s}}A$ in terms of the celestial symmetry parameters $\Aa{s}$ is canonically represented too.
The Noether charge associated to this transformation is denoted $\cqAas$, i.e. $\d{s}{\Aa{s}}\cdot=\poisson{\cqAas,\cdot}$, and we give its explicit expression in the next section.

To be more accurate about the notion of `non-perturbative' used here, if we formally introduce $g$ and $\bar g$ as small dimensionless parameters, then by rescaling $A\rightarrow gA$ and $\bA\rightarrow\bar g\bA$, our result is non perturbative in $g$ and at leading order in $\bar g$.\footnote{Recall that $\tQ_{-1}=-\bnews\propto\bar g$ while the EOM \eqref{EOMYM} or their dual \eqref{dualEOMYM} only introduce extra $A$'s.}
$g$ and $\bar g$ are a bookkeeping device and play the role of a complexification of the gauge coupling constant once we impose $g\bar g=\gYM^2$.

\section{Renormalized charge \label{secYM:RenormCharge}}

In \cite{Freidel:2023gue}, the charge renormalization procedure was necessary to get a finite action of the charge aspect in the limit $u\to-\infty$.
In \cite{Geiller:2024bgf}, it was then discussed at length (in the case of General Relativity, but YM would be the same) that even if the final formula for the action on phase space was correct, the actual expression of the renormalized charge aspect $\hq_s$ in terms of the original charges aspects $\tQ_s$ was subtler than the ansatz of \cite{Freidel:2023gue}.

Next, in \cite{Cresto:2024mne} (see also the work of \cite{Kmec:2024nmu} for a twistorial perspective), we proved that the charge renormalization procedure amounts to trading the Carrollian symmetry parameter $\a{s}$ for its celestial counterpart $\Aa{s}$.
The renormalization is thus no longer an iterative procedure to remove divergences.
It rather reflects the change of perspective between the Carrollian and the celestial descriptions of the symmetry.
The complicated field dependency necessary to construct the corner/celestial charge aspect $\hq_s(z,\bz)$ is then fully determined by the dual EOM for $\a{}$.

The systematic usage of the dual EOM furnishes an \textit{algorithmic, non-ambiguous and non-perturbative} (in $g$) way of defining the renormalized charges for any helicity.

On top of that, as it was pointed out in \cite{Freidel:2021dfs, Freidel:2021ytz, Geiller:2024bgf}, the finiteness of the action generated by $\hq_s$ comes hand in hand with the answer to the question ``In the absence of radiation, what is the conserved charge associated to $\tQ_s$?''.

Let us now see how the knowledge of the explicit solution $\bfa(\Aa{})\in\cS$ allows to construct the Noether charge $\cqAas$ associated to the celestial symmetry parameter $\Aa{s}$.

Turning words into equations, we define (cf. \eqref{masterChargeYM})
\begin{equation}
   \cqAas^u:= Q^u_{\bfa(\Aa{s})}=\sum_{n=0}^sQ^u_n\big[\a{n}(\Aa{s})\big]. \label{defcqasu}
\end{equation}
The following theorem then summarizes the aforementioned points.

\begin{tcolorbox}[colback=beige, colframe=argile]
\textbf{Theorem [Noether charge of spin $s$]}\\
The Noether charge $\cqAas$ associated to the celestial symmetry parameter $\Aa{s}$ of helicity $s$ is written as the following corner integral:
\begin{equation} \label{NoetherhatqsYM}
    \cqAas\equiv\cqAas^{-\infty} =\frac{2}{\gYM^2}\int_{\scri^+_-}\big\langle\hq_s,\Aa{s}\big\rangle_{\includegraphics[width=1.4mm]{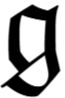}}= \lim_{u\to-\infty}\cq_s^{u}[\Aa{s}], \qquad s\geqslant 0, 
\end{equation}
with $\cqAas^u$ given by \eqref{defcqasu} and
\begin{equation}
    \hq_s(z,\bz)=\lim_{u\to -\infty}\tq_s(u,z,\bz).
\end{equation}
The renormalized charge aspect $\tq_s$ satisfies $\pa_u\tq_s=0$ in the absence of left-handed radiation $\bnews=0$.
\end{tcolorbox}

\paragraph{Proof:}
We construct $\tq_s$ systematically in the appendix \ref{AppYM:renormcharge}, cf. \eqref{deftqsYM}, where we also give explicit expressions for $s=0,1,2$ in \eqref{renormchargeYM}.
To show that $\tq_s$ is conserved in the absence of left-handed radiation, notice that (cf. \eqref{Qadot})
\begin{equation}
    \pa_u \cqAas^u =\frac{2}{\gYM^2}\int_S\big\langle\bnews,\delta_{\bfa{}(\Aa{s})}A\big\rangle_{\includegraphics[width=1.4mm]{test2}}=0 \qquad\textrm{if}\qquad\bnews=0,
\end{equation}
while we also have that
\begin{equation}
    \pa_u \cqAas^u =\frac{2}{\gYM^2}\pa_u\left(\int_S\big\langle\tq_s,\Aa{s}\big\rangle_{\includegraphics[width=1.4mm]{test2}}\right)= \frac{2}{\gYM^2}\int_S \big\langle\pa_u\tq_s,\Aa{s}\big\rangle_{\includegraphics[width=1.4mm]{test2}}.
\end{equation}
This concludes the proof.

\paragraph{Remark:}
In a non-radiative strip of $\scri$, where $\news=0$ for $u\in[0,u_0]$, we get that
\begin{equation}\label{deftqs0radYM}
    \boxed{\tq_s=\sum_{k=0}^s\frac{(-u)^k}{k!}\Dcal^k\tQ_{s-k}}.
\end{equation}
The proof is straightforward since for $u\in[0,u_0]$,
\begin{align}
    \cq^u_s[\Aa{s}]\overset{\eqref{defcqasu}}{=}\sum_{n=0}^sQ^u_n\big[\a{n}(\Aa{s})\big] &\overset{\eqref{solAlphanAs}}{=}\sum_{n=0}^sQ^u_n\big[\big(\pui\Dcal\big)^{s-n}\Aa{s}\big] \nn\\
    &\overset{F=0}{=} \sum_{n=0}^s\frac{u^{s-n}}{(s-n)!}Q^u_n\big[\Dcal^{s-n}\Aa{s}\big]\overset{\eqref{QDDQ}}{=} \sum_{n=0}^s\frac{(-u)^{s-n}}{(s-n)!}\Dcal^{s-n}Q^u_n\big[\Aa{s}\big],
\end{align}
where to go from the first to the second line, we used that $F=0$ so that $\big(\pui\Dcal\big)^{s-n}\Aa{s} =\Dcal^{s-n}\pui[(s-n)]\Aa{s}=\frac{u^{s-n}}{(s-n)!}\Dcal^{s-n}\Aa{s}$.
We then extract \eqref{deftqs0radYM} upon changing $s-n\to k$.
The reader can check that this renormalized charge aspect is conserved in the strip if we also assume that $\bnews=0$.

\section{Action on the gauge potential \label{secYM:ActionOnA}}

As an application of the general formalism presented so far, we compute the soft and hard actions of the higher spin charges $\cqAas$ on the asymptotic gauge potential $A$.

The Noether charge $\cqAas$, or equivalently the charge aspect $\hq_s$, depends linearly on $\bnews$ and has a polynomial dependence\footnote{The coefficients of the polynomial are in general differential operators.} on $A$. 
More precisely, for $s\geqslant -1$, we can decompose $\hq_s$ as
\be 
\hq_s =\sum_{k=0}^{s+1} \q{s}{k},
\ee 
where $\q{s}{k}$ is homogeneous of degree $k$ in $A$ and linear in $\bnews$.\footnote{Recall that this is equivalent to the expansion in terms of the coupling constant $g$ (while all charges are at leading order in $\bar g$).}
The charge aspect $\q{s}{0}\equiv\soft{\hq_s}$ is the soft charge while $\q{s}{1}\equiv\hard{\hq_s}$ is the hard charge and $\sum_{k=2}^{s+1} \q{s}{k}$ is the super-hard contribution.
Here we compute the soft and hard actions for any spin. 
We let for future work the study of the super-hard contribution.
However, we underline that the algebra generated by $\Qa$ does close \textit{without} relying on any soft or hard truncation; this is the main point of this paper.

We classify the action as a function of the helicity by defining 
\begin{equation}
    \d{s}{\Aa{s}}A :=\delta_{\bfa(\Aa{s})}A= \poisson{\cqAas,A}, \label{defdeltaAs}
\end{equation}
where $\bfa(\Aa{s})$ is the solution \eqref{solAlphanAs}.
The superscript $\d{s}{}$ refers to the helicity of the charge the transformation is associated with.
This notation will come handy in the following.
Moreover, since the fields do not diverge at $|u|\to\infty$ due to the Schwartz fall-off condition, the operation $\pa_u\pui$ is the identity and we can conveniently write
\begin{equation}
    -\d{s}{\Aa{s}}A=\Dcal\a0(\Aa{s})=\Dcal \big(\pui\Dcal\big)^s\Aa{s}=\pa_u\big(\pui\Dcal\big)^{s+1}\Aa{s}.
\end{equation}
We also denote\footnote{$\soft{\cq_{\Aa{s}}} =\frac{2}{\gYM^2}\int_S\big\langle\soft{\hq_s}, \Aa{s}\big\rangle_{\includegraphics[width=1.4mm]{test2}}$ and similarly for $\hard{\cq_{\Aa{s}}}$.}
\begin{equation}
    \poisson{\soft{\cqAas},A} =\Sd{s}{\Aa{s}} A\qquad\textrm{and}\qquad \poisson{\hard{\cqAas},A}=  \Hd{s}{\Aa{s}}A.
\end{equation}
Before stating the results, we define yet another piece of notation, namely 
\begin{equation}
    \td{s}{\Aa{s}}A\equiv\deg^u_0\big(\d{s}{\Aa{s}}A\big),
\end{equation}
where $\deg^u_0(O)$ stands for the coefficient of the term $u^0$ in the expression $O$ (when $O$ is a polynomial in $u$).

\begin{tcolorbox}[colback=beige, colframe=argile]
\textbf{Lemma [Soft action]}\\
The soft action for arbitrary spin $s\in\N$ is given by
\begin{equation}
    \Sd{s}{\Aa{s}}A=-\frac{u^s}{s!} D^{s+1}\Aa{s}. \label{SoftActionYM}
\end{equation}
\end{tcolorbox}

\paragraph{Proof:} 
It is direct since the soft part of $\Dcal^{s+1}$ is $D^{s+1}$, so that
\begin{equation}
    \Sd{s}{\Aa{s}} A=-\soft{\Big(\pa_u\big( \pui\Dcal\big)^{s+1}\Aa{s}\Big)}=-\pa_u\big(\pui D\big)^{s+1}\Aa{s}=-\frac{u^s}{s!}D^{s+1}\Aa{s}.
\end{equation}
Notice that $\Sd{s}{\Aa{s}}A=\frac{u^s}{s!}\Std{0}{D^s\Aa{s}}A$, where $\Std{0}{\Aa0}A=-D\Aa0$.
This means that the soft transformation of spin $s$, parametrized by the tensor $\Aa{s}$, is simply the soft part of a super-phase rotation parametrized by the tensor $D^s\Aa{s}$ (times the $u$ dependence $\frac{u^s}{s!}$).

\begin{tcolorbox}[colback=beige, colframe=argile]
\textbf{Lemma [Hard action]}\\
The hard action for arbitrary spin $s\in\N$ takes the form
\begin{subequations}
\label{YMHardAction}
\begin{equation}
    \Hd{s}{\Aa{s}}A=\sum_{p=0}^s\frac{u^{s-p}}{(s-p)!}\Htd{p}{D^{s-p}\Aa{s}}A,
\end{equation}
where
\begin{equation}
    \Htd{0}{\Aa0}A=-[A,\Aa0]_\g \quad~\textrm{and }\quad \Htd{p}{\Aa{p}}A=-D\big[\pui[p]D^{p-1}A,\Aa{p}\big]_\g,\quad p\geq 1.
\end{equation}
\end{subequations}
\end{tcolorbox}

\paragraph{Proof:}
We have to consider 
\begin{equation}\label{puiDcalHard}
    \Hd{s}{\Aa{s}} A=-\hard{\Big(\pa_u\big( \pui\Dcal\big)^{s+1}\Aa{s}\Big)},
\end{equation}
where $\big( \pui\Dcal\big)^{s+1}$ is given by \eqref{puiDcalkYM}.
In \eqref{puiDcalkYM}, the sum over $\l$ precisely counts the number of gauge potential, i.e. the expansion in $g^\l$.
The hard part, namely the part proportional to $g$, is thus the term $\l=1$.
We present the proof of how to relate \eqref{puiDcalHard} to \eqref{YMHardAction} in appendix \ref{App:HardActionYM}.

A similar computation was carried out in \cite{Freidel:2023gue} using a discrete basis of modes for $A$ (see also \cite{Freidel:2022skz}).
Since it was not obvious from \cite{Freidel:2023gue}, we find remarkable that the hard action specific to the degree $p$, namely $\Htd{p}{\Aa{p}}A$, reduces to a total derivative (for $p\geq 1$).
More precisely, our formula \eqref{YMHardAction} is the same as equation (50) in \cite{Freidel:2023gue}.
The proof necessitates some manipulations so we report it in appendix \ref{AppYM:comparison}.
The point is that \eqref{YMHardAction} now appears as a consequence of the general, non-perturbative action of the Noether charge $\Qa$, in a much simpler way than the perturbative treatment known so far.

\section{Wedge sub-algebras of $\cS$ \label{secYM:wedge}}

In this section, we discuss (wedge) sub-algebras present in $\cS$ or its projection at a cut of $\scri$. 

\subsection{$\Wcal_A(\scri)$-algebra}

The Lie algebroid $\sfS$-bracket $\lbr\cdot\,,\cdot\rbr$ reduces to the Lie algebra bracket $[\cdot\,,\cdot]$ when $\da\ap{}\propto\da A\equiv 0$.
This imposes a set of constraints on the symmetry parameters $\a{}$.
Indeed, by definition $\da A=-\Dcal\a0$ is set to 0.
Then, since $\da A=0$ implies $\pa_u(\da A)=\da F=0$, and that the commutator of $\pa_u$ with $\Dcal$ is just
\begin{equation}
    [\pa_u,\Dcal]\,\cdot=[F,\cdot\,]_\g,
\end{equation}
we also have that
\begin{equation}\label{wedgeStart}
    0=-\da F=\pa_u(\Dcal\a0)=\Dcal (\pa_u\a0)+[F,\a0]_\g=\Dcal^2\a1+[F,\a0]_\g.
\end{equation}
Iterating this reasoning, the $s^{\rm{th}}$-time derivative of $\da A$ then generates a condition on $\Dcal^{s+1}\a{s}$.
To summarize, the first 3 conditions are
\bs
\begin{align}
    \Dcal\a0 &=0, \\
    \Dcal^2\a1 &=-[F,\a0]_\g, \\
    \Dcal^3\a2 &=-[\Dcal F,\a1]_\g -3[F,\Dcal\a1]_\g -[\pa_u F,\a0]_\g.
\end{align}
\es

We thus define the covariant wedge space as follows.
\begin{tcolorbox}[colback=beige, colframe=argile]
\textbf{Definition [Covariant wedge space on $\scri$]}\\
\begin{equation}
    \W_A(\scri):=\Big\{\,\a{}\in\sfS~ \big|~\pa_u^s\big(\Dcal\a0\big)=0,\, \forall \,s\in\N\,\Big\}.
\end{equation}
\end{tcolorbox}

\begin{tcolorbox}[colback=beige, colframe=argile]
\textbf{Theorem [Covariant wedge algebra on $\scri$]}\\
$\Wcal_A(\scri)\equiv\big(\W_A(\scri),[\cdot\,,\cdot]\big)$ is a Lie algebra over $\scri$.
\end{tcolorbox}

\paragraph{Proof:}
Since $\da A=0=\dap A$ implies that $\pa_u[\a{},\ap{}]=\Dcal[\a{},\ap{}]$, cf. App.\,\ref{App:dualEOMYM}, we are allowed to evaluate $\pa_u^s(\da A)$ for $\a{}\to[\a{},\ap{}]$. 
The $s^{\rm{th}}$-time derivative then generates a condition on $\Dcal^{s+1}\a{s}$ which is automatically preserved by the bracket.
In brief,
\begin{equation}
    \da A=0\quad\Longrightarrow\quad \Big(\da\big(\pa_u^s A\big)=0\quad\&\quad\delta_{[\a{},\ap{}]}\big(\pa_u^s A\big)=0\Big),\quad\forall \,s\in\N.
\end{equation}

Even if we let a detailed analysis of $\Wcal_A(\scri)$ for another work, we can say that it encompasses the symmetry transformations that leave the asymptotic gauge field $A$ invariant.
It is thus the symmetry algebra of such an asymptotic solution.

\subsection{$\Wcal_A(S)$-algebra \label{subsecYM:wedgeAlgebraOnS}}

We now focus on the particular case of a non-radiative cut of $\scri$, for which the analysis simplifies greatly.

First of all, notice that the $\cS$-algebroid over $\scri$ can be projected at any cut. 
For definiteness and to match with the solution of the dual EOM discussed in section \ref{secYM:solEOM}, we always consider the cut $S=\{u=0\}\subset\scri$.
Therefore, there exists a Lie algebroid bracket $\lbr\cdot\,,\cdot\rbr$ over $S$ which is naturally given by the evaluation of its Carrollian version at $u=0$, namely\footnote{Of course $[\Aa{},\Aap{}]_s=\sum_{n=0}^s[\Aa{n},\Aap{s-n}]_\g$.}
\begin{equation}
    \lbr\Aa{},\Aap{}\rbr:=\lbr\a{},\ap{}\rbr \big|_{u=0}=[\Aa{},\Aap{}]+\dAp\Aa{}-\dA\Aap{}.
\end{equation}

Second of all, if $\Dcal\Aa0=0=\Dcal\Aap0$, then the action of the anchor map vanishes and $\lbr\Aa{},\Aap{}\rbr$ reduces to the Lie algebra bracket $[\Aa{},\Aap{}]$.
When writing $\Dcal\Aa0$, it is understood that the gauge covariant derivative is evaluated at the cut too. 
In other words $\Dcal\Aa{s}:=D\Aa{s}+[\Ac0,\Aa{s}]_\g$, where $\Ac0:=A\big|_{u=0}$.
With that notation, $\dA \Ac0=\big(\da A\big)\big|_{u=0}=-\Dcal\Aa0$.

Third of all, if we consider a fully non-radiative cut, by which we mean that $\Ac{n}:=\big(\pa_u^n A\big)\big|_{u=0} \equiv 0$ for $n\geq 1$, then the symmetry of the cut has to preserve these conditions.
Using the calculation \eqref{wedgeStart}, we get
\begin{align}
    -\dA\Ac1=-\big(\da F\big)\big|_{u=0} =\Dcal^2\Aa1+\big[\Ac1,\Aa0 \big]_\g.
\end{align}
We thus see that in order to preserve the condition $\Ac1=0$, we need to take $\Dcal^2\Aa1=0$.
Iterating this computation, we deduce that
\begin{equation}
    -\dA\Ac{n}=\big(\pa_u^n(\Dcal\a0)\big) \big|_{u=0}=\Dcal^{n+1}\Aa{n},\qquad\textrm{if}\quad \Ac{n}=0,\,n\geqslant 1.
\end{equation}
Hence preserving the non-radiative nature of the cut under consideration amounts to imposing the \textit{covariant wedge} condition $\Dcal^{s+1}\Aa{s}=0$, $s\in\N$.

To summarize our findings, we define the $\W_A(S)$-space and state the associated theorem.

\begin{tcolorbox}[colback=beige, colframe=argile]
\textbf{Definition [Covariant wedge space on $S$]}\\
\begin{equation}
    \W_A(S):=\Big\{\,\Aa{}\in\V(S,\g)~ \big|~\Dcal^{s+1}\Aa{s}=0,\, \forall\, s\in\N\,\Big\}.
\end{equation}
\end{tcolorbox}

\begin{tcolorbox}[colback=beige, colframe=argile]
\textbf{Theorem [Covariant wedge algebra on $S$]}\\
$\Wcal_A(S)\equiv\big(\W_A(S),[\cdot\,,\cdot]\big)$ is a Lie algebra over $S$.
\end{tcolorbox}

\paragraph{Proof:}
From the intrinsic perspective of the sphere, i.e. not using the projection from $\scri$ of $\Wcal_A(\scri)$, the reason why the bracket preserves the covariant wedge condition is that 
\begin{equation}
    \Dcal^{s+1}[\Aa{},\Aap{}]_s=\sum_{n=0}^s \sum_{k=0}^{s+1}\binom{s+1}{k}\big[ \Dcal^k\Aa{n},\Dcal^{s+1-k}\Aap{s-n}\big]_\g=0 \label{proofCovWedgeYM}
\end{equation}
since $\Dcal^k\Aa{n}\equiv 0$ for $k\geqslant n+1$ and $\Dcal^{s+1-k}\Aap{s-n}\equiv 0$ for $k\leqslant n$.
Hence $[\Aa{},\Aap{}]\in\Wcal_A(S)$ if $\Aa{},\Aap{}\in\Wcal_A(S)$.

\paragraph{Remark:}
As we can see from the proof \eqref{proofCovWedgeYM}, the $\cS$-algebroid admits a collection of sub-algebras $\WS{n}$ which are organized into a filtration
\begin{equation}
    \WS{}\subset\ldots\subset\WS{n+1} \subset\WS{n} \subset\ldots\subset\WS0, 
\end{equation}
where\footnote{We use the same letter $\cS$ to designate both the algebroid over $\scri$ and the one over $S$.
Since we use $\a{}$ and $\Aa{}$ for the respective variables, the context always makes it clear which algebroid we are referring to.}${}^,$\footnote{The condition $\Dcal\Aa0$ is always satisfied, so that the anchor $\dA$ always vanishes and $\WS{n}$ is a Lie algebra for any $n\in\N$.}
\begin{equation}
    \WS{n}:=\Big\{\,\Aa{}\in\cS~ \big|~\Dcal^{s+1}\Aa{s}=0,\, 0\leqslant s\leqslant n\,\Big\}\subset\cS
\end{equation}
and $\WS{}=\lim_{n\to\infty}\WS{n}$.

\paragraph{Remark:}
The discussion of this subsection is alike the one for GR \cite{Cresto:2024fhd, Cresto:2024mne} where $\Wcal_A(S)$ here corresponds to $\Wcal_\sigma(S)$ there.
In both cases, the covariant wedge comes about as the non-radiative restriction of the projection of $\Wcal_A(\scri)$ (or $\Wcal_C(\scri)$ in GR) at a cut.
In the gravitational study, there is however another layer where the covariant wedge appears intrinsically from the sphere \cite{Cresto:2024fhd} (i.e. not from the projection $\scri\to S$).
The reason being that the deformed $\sigma$-bracket is a Lie bracket only \textit{inside} $\Wcal_\sigma(S)$.
The Jacobi identity is otherwise violated, with the violation proportional to $\dT\sigma$, where $\dT\sigma$ is the gravitational analog of $\dA \Ac0$.
The bracket $[\cdot\,,\cdot]$ (equivalently $[\cdot\,,\cdot]_\g$) does not undergo any deformation in the gauge theory case and is thus a genuine Lie bracket irrespectively of any conditions on $\Aa{s}$.
Remarkably, in the gravitational case, $\dT\sigma=0$ \textit{amounts} to the covariant wedge onto the sphere; namely it leads to conditions on all $\T{s},\,s\geq 0$.
For YM, we just saw that $\Dcal\Aa0=0$ does not impose any constraint on $\Aa1$.\footnote{This behavior can be traced back to the fact that $[\cdot\,,\cdot]$ is of degree 0 while its gravitational counterpart, namely the $\sigma$-bracket $[\cdot\,,\cdot]^\sigma$ is of degree $-1$.} 
In that sense, $\Wcal_A(S)$ appears \textit{solely} as the projection of $\Wcal_A(\scri)$ at a non-radiative cut.

\paragraph{Remark:}
From the general result about the Noether representation in Sec.\,\ref{secYM:Noether}, we infer that the covariant wedge algebra is represented canonically, so that
\begin{equation}
    \boxed{\,\poisson{\cq_{ \text{\scalebox{1.24}{$\Aa{}$}}},\cq_{ \text{\scalebox{1.24}{$\Aap{}$}}}}=- \cq_{\text{\scalebox{1.24}{$[\Aa{},\Aap{}]$}}{}^{\includegraphics[width=1.4mm]{test2}}}\,}.
\end{equation}

\section{Relation to twistor theory \label{secYM:twistor}}

As in \cite{Cresto:2024mne}, we present an alternative description of the Carrollian symmetry parameters $\a{}$ by trading the spin degree for a continuous dimension labeled by a spin 1 variable $q \in \Ccar{0,1}$.
This allows to promote the series $\a{}=(\a{s})_{s\in \N}$ to a $\g$-valued holomorphic function $\fa$ of $q$\footnote{The Bondi coordinates dependence is implicit.} 
\be
\fa(q)=\sum_{s=0}^{\infty} \a{s}q^s \in \tV_0.
\ee 
The graded vector $\a{}$ and the function $\fa$ are two representations of the same abstract vector in $\tV(\scri,\g)$.
The dual EOM are represented on these functions as the functional
\begin{equation}
    \sE_{\fa}(q):= \sum_{s=0}^\infty \big(\pa_u\a{s}-(\Dcal\a{})_s\big)q^s.
\end{equation}
Moreover we see that the function representation of the bracket $[\a{},\ap{}]$ is the Lie algebra $\g$-bracket of the function representation of $\a{},\ap{}$.
Indeed,
\begin{align}
    [\fa, \fa'](q):=\sum_{s=0}^\infty[\a{}, \ap{}]_sq^s =
    \sum_{s=0}^\infty\sum_{n=0}^s[\a{n},\ap{s-n}]_\g q^s &=\sum_{n=0}^\infty \sum_{s=n}^\infty\big[\a{n}q^n,\ap{s-n}q^{s-n}\big]_\g \cr &=\big[\fa(q),\fa'(q)\big]_\g.
\end{align}
Finally, introducing the covariant derivative 
\be 
\nabla := q\pa_u -D-\ada,
\ee
we obtain that 
\begin{align}
    \nabla \fa(q) =\sum_{s=0}^\infty \big(\pa_u\a{s}- 
    \Dcal\a{s+1}\big) q^{s+1}
    - \Dcal\a0= q\E_{\fa}(q)+\da A.\label{bNI}
\end{align}
The condition $\a{}\in\sfS$, which imposes $\E_{\fa}=0$, simply reads $\pa_q \nabla \fa=0$.
It shows that when $\a{}\in \cS$, then $\nabla\fa$ is independent of $q$ and
\begin{equation}
\da A\equiv\dah A=\nabla \fa.
\end{equation}
Here we see that the dual EOM appear when requiring that the infinitesimal field transformation $\dah A$ be independent of the coordinate $q$.
This is natural since the asymptotic gauge field $A$ is by construction only a function of $(u,z,\bz)$.

It is customary to get field dependent symmetry parameters after a partial gauge fixing procedure.
In our case, we can tie this $q$-independence of $\dah A$ to the analogous twistor gravitational study of \cite{Kmec:2024nmu}, where $q$ is precisely the twistor coordinate in the fiber above $\scri$ \cite{Adamo:2021lrv}.\footnote{See also \cite{Cresto:2024mne} for a discussion about the interpretation of $q$ in the context of Newman's good cut equation \cite{newman_heaven_1976, Adamo:2010ey} and as a parametrization of an Ehresmann connection in the Carrollian context \cite{Mars:1993mj, Freidel:2024emv}.}
Indeed, assuming that the twistor gauge potential comes from the lift of a $\scri$ gauge potential, the former must be independent of the fiber coordinate $q$.
From a twistor space perspective, this can be seen as a gauge fixing.
With that insight, $\dah A$ should result from a twistor gauge transformation when studying self-dual YM theory in twistor space.

\section*{Acknowledgments}
\addcontentsline{toc}{section}{Acknowledgments}

I thank Laurent Freidel for the support along this project.
I also thank Romain Ruzziconi, Lionel Mason and Adam Kmec, together with Daniele Pranzetti, Shreyansh Agrawal, Laura Donnay and Panagiotis Charalambous, for discussions during my visit respectively at the Mathematical Institute in Oxford and at SISSA in Trieste.
I thank as well Marc Geiller together with Tom Wetzstein and Laurent Baulieu for the fruitful interactions while visiting ENS Lyon and LPTHE Paris respectively.
Thanks to Hank Chen for some discussions during the early stages of this project.

\section*{Funding and Competing interests}
\addcontentsline{toc}{section}{Funding and Competing interests}

The authors declare they have no financial or conflict of interests.
Research at Perimeter Institute is supported by the Government of Canada through the Department of Innovation, Science and Economic Development and by the Province of Ontario through the Ministry of Colleges and Universities. This work was supported by the Simons Collaboration on Celestial Holography.

\section*{Data Availability Statement}
\addcontentsline{toc}{section}{Data Availability Statement}

No datasets were generated or analyzed during the current study.

\newpage
\appendix
\section{Proof of the Jacobi identity \label{App:JacobiYM}}

To prove that the $\sfS$-bracket $\lbr\cdot,\cdot\rbr$ is indeed a Lie (algebroid) bracket, we check that it satisfies the Jacobi identity.
First notice that $\big\lbr\a{},\lbr\ap{},\app{}\rbr\big\rbr_s$ splits into 3 contributions:
\begin{subequations}
\begin{align}
    \big\lbr\a{},\lbr\ap{},\app{}\rbr \big\rbr_s &=-\da\lbr\ap{},\app{}\rbr_s+ \delta_{\lbr\ap{},\app{}\rbr}\a{s} +\sum_{n=0}^s\big[\a{n},\lbr\ap{},\app{}\rbr_{s-n}\big]_\g \nn\\
    &=\da\dap\app{s}-\da\dapp\ap{s}+ \delta_{\lbr\ap{},\app{}\rbr}\a{s}\label{line1}\\
    &-\sum_{n=0}^s\Big(\big[\da\ap{n},\app{s-n}\big]_\g+\big[\ap{n},\da\app{s-n}\big]_\g\Big)-\sum_{n=0}^s\big[\a{n},\dap\app{s-n}-\dapp\ap{s-n}\big]_\g \label{line2}\\
    &+\sum_{n=0}^s\sum_{k=0}^{s-n}\big[\a{n},[\ap{k},\app{s-n-k}]_\g\big]_\g. \label{line3}
\end{align}
\end{subequations}
Taking the cyclic permutation,\footnote{We use the notation $\cyc$ to denote an equality valid upon adding the cyclic permutation of the terms on the left and right hand sides.} we get that the first line vanishes,
\begin{equation}
    \eqref{line1}\cyc \da\dap\app{s}-\dap\da\app{s}+ \delta_{\lbr\a{},\ap{}\rbr}\app{s}= 0,
\end{equation}
where we used the morphism property \eqref{SalgebraMorphism}.
Similarly, all terms of \eqref{line2} cancel once we consider its cyclic permutation.
Indeed,
\begin{align}
    -\eqref{line2}&\cyc \sum_{n=0}^s\Big(\big[\da\ap{n},\app{s-n}\big]_\g+\big[\ap{n},\da\app{s-n}\big]_\g\Big)+\sum_{n=0}^s\big[\app{n},\da\ap{s-n}\big]_\g -\sum_{n=0}^s\big[\ap{n},\da\app{s-n}\big]_\g \nn\\
    &\cyc \sum_{n=0}^s\big[\da\ap{n},\app{s-n}\big]_\g+\sum_{n=0}^s\big[\app{s-n},\da\ap{n}\big]_\g = 0,
\end{align}
where to get the second equality, we changed $n\to s-n$ in the third term of the first line.
The contribution \eqref{line3} vanishes thanks to the Jacobi identity of the Lie algebra bracket $[\cdot\,,\cdot]_\g$.\footnote{Simply notice that $\eqref{line3}=\sum_{a+b+c=s}\big[\a{a},[\ap{b},\app{c}]_\g\big]_\g$ and that the sum is now invariant under permutation of $a,b,c$.}
We thus infer that $\big\lbr\a{},\lbr\ap{},\app{}\rbr \big\rbr\cyc 0$.

\section{Closure of the $\sfS$-bracket and Leibniz anomalies \label{App:dualEOMYM}}

Recall that we define the Leibniz rule anomaly for any bracket $[\cdot\,,\cdot]$ and differential operator $\mathscr{D}$ as
\begin{equation}
    \cA\big([\cdot\,,\cdot],\mathscr{D} \big)=\mathscr{D}[\cdot\,,\cdot]-[\mathscr{D}\cdot\,,\cdot]-[\cdot\,,\mathscr{D}\cdot].
\end{equation}
We already proved in \eqref{DcalLeibniz} that
\begin{equation}
    \cA\big([\cdot\,,\cdot]_\g,\Dcal\big)=0.
\end{equation}
However, notice that
\begin{align}
    \big(\Dcal[\a{},\ap{}]\big)_{s} &= \Dcal[\a{},\ap{}]_{s+1}=\sum_{n=0}^{s+1}\Big(\big[(\Dcal\a{})_{n-1},\ap{s+1-n} \big]_\g+\big[\a{n},(\Dcal\ap{})_{s-n}\big]_\g\Big) \\
    &=\big[(\Dcal\a{})_{-1},\ap{s+1}\big]_\g \!+\!\sum_{n=0}^{s}\big[(\Dcal\a{})_n,\ap{s-n} \big]_\g\!+\big[\a{s+1},(\Dcal\ap{})_{-1}\big]_\g\!+\!\sum_{n=0}^{s} \big[\a{n},(\Dcal\ap{})_{s-n}\big]_\g, \nn
\end{align}
while
\vspace{-0.2cm}
\begin{equation}
    \big[\Dcal\a{},\ap{}\big]_s+\big[\a{}, \Dcal\ap{}\big]_s= \sum_{n=0}^{s}\big[(\Dcal\a{})_n,\ap{s-n} \big]_\g+\sum_{n=0}^{s} \big[\a{n},(\Dcal\ap{})_{s-n}\big]_\g.
\end{equation}
Consequently, using $(\Dcal\a{})_{-1}=\Dcal\a0=-\da A$, we get that\footnote{Note that this anomaly vanishes inside $\Wcal_A(\scri)$ (or $\Wcal_A(S)$ if we trade $\a{}$ for $\Aa{}$), see Sec.\,\ref{secYM:wedge}.}
\begin{equation}
    \boxed{\cA_s\big([\a{},\ap{}],\Dcal\big)= \big[\dap A,\a{s+1}\big]_\g-\big[\da A,\ap{s+1}\big]_\g}.
\end{equation}
Next, using that
\begin{align}
    \Dcal\big(\dap\a{s+1}-\da\ap{s+1}\big) &=\dap(\Dcal\a{})_s-\da(\Dcal\ap{})_s-\big[\dap A,\a{s+1}\big]_\g+\big[\da A,\ap{s+1}\big]_\g \nn\\
    &=\dap(\Dcal\a{})_s-\da(\Dcal\ap{})_s-\cA_s\big([\a{},\ap{}],\Dcal\big),
\end{align}
we infer that
\begin{align}
    \big(\Dcal\lbr\a{},\ap{}\rbr\big)_s &=\big(\Dcal[\a{},\ap{}]\big)_s+\Dcal\big(\dap\a{s+1}-\da\ap{s+1}\big) \cr
    &=\Big(\big[\Dcal \a{}, \ap{}\big]_s+\dap(\Dcal\a{})_s\Big)-\a{}\leftrightarrow\ap{}.
\end{align}
Since $\big\lbr\Dcal\a{},\ap{}\big\rbr_s= \big[\Dcal \a{}, \ap{}\big]_s+\dap (\Dcal\a{})_s-\delta_{\Dcal \a{}}\ap{s}$, we finally obtain that
\begin{equation}
    \cA\big(\lbr\a{},\ap{}\rbr, \Dcal\big)=\delta_{\Dcal \a{}}\ap{}-\delta_{\Dcal \ap{}}\a{}.
\end{equation}
We can also compute similar anomalies for $\pa_u$:
\begin{equation}
     \cA\big([\cdot\,,\cdot]_\g,\pa_u\big) =0\qquad\textrm{and}\qquad \cA\big([\cdot\,,\cdot],\pa_u\big)=0,
\end{equation}
while
\begin{equation}
    \cA\big(\lbr\a{},\ap{}\rbr, \pa_u\big)=\delta_{\pa_u\a{}}\ap{}-\delta_{\pa_u\ap{}}\a{}.
\end{equation}

\section{Proof of Theorem \hyperref[NoetherhatqsYM]{[Noether charge at spin $s$]} \label{AppYM:renormcharge}}

The goal of this demonstration is to recast the Noether charge $\Qa$, when $\a{}=\bfa(\Aa{s})$, as the integral over the (possibly punctured) sphere $S$ of a certain renormalized charge aspect $\tq_s(u,z,\bz)$ smeared against the symmetry parameter $\Aa{s}$ that defines the $\bfa(\Aa{s})$.

Let us start with an example for the lowest spin-weights.
Take $\alpha\in\cS^2$, which implies that $\Aa{}=(\Aa0,\Aa1,\Aa2,0,\ldots)$, then
\bs \label{sola210}
\begin{align}
    \a2(\Aa{}) &=\Aa2, \\
    \a1(\Aa{}) &=uD\Aa2+\Aa1+\pui[1][A,\Aa2]_\g, \\
    \a0(\Aa{}) &=\frac{u^2}{2}D^2\Aa2+uD\Aa1+ \Aa0+\pui[2]D[A,\Aa2]_\g \\
    &+\pui\big[A,uD\Aa2+ \Aa1\big]_\g+\pui\big[A,\pui[1][A,\Aa2]_\g\big]_\g. \nn
\end{align}
\es
Consider successively $\a{}=\bfa(\Aa{s})$ for $s=0,1,2$.
Using the explicit computation \eqref{sola210}, we deduce that\footnote{We use the trace for convenience of writing, which introduces a sign, cf. footnote \ref{footPairg}.}${}^,$\footnote{Notice that some steps need care. 
For instance, $\Tr\Big(\tQ_0\pui\big[A,uD \Aa2\big]_\g\Big)=\Tr\Big(\tQ_0\big[\pui(uA),D \Aa2\big]_\g\Big)=\Tr\Big(\big[\tQ_0, \pui(uA)\big]_\g D\Aa2\Big)=\Tr\Big(-D\big[\tQ_0, \pui(uA)\big]_\g\Aa2\Big)$ up to a total derivative that drops upon integration.}
\bs
\begin{align}
    \frac{\gYM^2}{-2}\cq^u_{\text{\scalebox{1.24}{$\Aa0$}}} &=\int_S\Tr\big(\tQ_0\Aa0\big), \\
    \frac{\gYM^2}{-2}\cq^u_{\text{\scalebox{1.24}{$\Aa1$}}} &=\int_S\Tr\left(\tQ_0 \big(uD\Aa1+\pui[1][A,\Aa1]_\g\big)+\tQ_1\Aa1\right) \nn\\
    &=\int_S\Tr\bigg(\Big(-uD\tQ_0+\big[\tQ_0,\pui A\big]_\g+\tQ_1\Big)\Aa1\bigg), \\
    \frac{\gYM^2}{-2}\cq^u_{\text{\scalebox{1.24}{$\Aa2$}}} &=\int_S\Tr\bigg(\tQ_0\left(\frac{u^2}{2}D^2\Aa2+\pui[2]D[A,\Aa2]_\g+\pui\big[A,uD\Aa2\big]_\g+\pui\big[A,\pui[1][A,\Aa2]_\g\big]_\g\right) \nn\\
    &\qquad\quad\, +\tQ_1\big(uD\Aa2+\pui[1][A,\Aa2]_\g\big)+\tQ_2\Aa2\bigg) \nn\\
    &= \int_S\Tr\bigg(\bigg(\frac{u^2}{2}D^2\tQ_0-\big[D\tQ_0,\pui[2]A\big]_\g -D\big[\tQ_0,\pui(uA)\big]_\g+\Big[\big[\tQ_0, \pui \big(A\big]_\g,\pui A\big)\Big]_\g \nn\\
    &\qquad\qquad -uD\tQ_1+\big[\tQ_1,\pui A\big]_\g+\tQ_2\bigg)\Aa2\bigg).
\end{align}
\es
We can thus identify the renormalized charge aspects\footnote{Be mindful of the peculiar set of parenthesis when one starts to have nested commutators in $\tq_2$.}
\vspace{-0.4cm}
\begin{adjustwidth}{-0.25cm}{-0.1cm}
\bs
\label{renormchargeYM}
\begin{align}
\tq_0 &\equiv q_0, \\
\tq_1 &\equiv q_1+\big[\tQ_0,\pui A\big]_\g, \\
\tq_2 &\equiv q_2-\!\big[D\tQ_0,\pui[2]A\big]_\g -D\big[\tQ_0,\pui(uA)\big]_\g\! +\Big[\big[\tQ_0, \pui \big(A\big]_\g,\pui A\big)\Big]_\g\!+\big[\tQ_1,\pui A\big]_\g, 
\end{align}
\es
\end{adjustwidth}
such that $\cqAas^u=\cq_s^u[\Aa{s}]$, $s=0,1,2$, where 
\be 
q_s =\sum_{n=0}^s \frac{(-u)^n}{n!} D^n \tQ_{s-n}. \label{defqsYM}
\ee 
The reader can check that $\pa_u\tq_s$ only contains terms that involve $\bnews$ so that $\tq_s$ is conserved when no left-handed radiation is present.

Let us mention that the expression for $\cq^u_{\text{\scalebox{1.24}{$\Aa1$}}}$ in the abelian case first appeared in \cite{Lysov:2014csa}.
In that case we have $\cq^u_{\text{\scalebox{1.24}{$\Aa1$}}}=\frac{2}{e^2}\int_S\big(\tQ_0(u\pa_z\Aa1)+\tQ_1\Aa1\big)$, with $e$ the electric charge, which matches with (4.9) of \cite{Lysov:2014csa} since $\lim_{r\to\infty}\big(r^2\mathcal{F}_{ru}\gamma_{z\bz}+\mathcal{F}_{\bz z}\big)=2\tQ_0$ and $\lim_{r\to\infty}\big(r^2\mathcal{F }_{rz}\big)=\tQ_1$, while $\Aa1=Y^z=Y_{\bz}$.\footnote{Just for simplicity, we wrote the expressions on the celestial plane where $D\to\pa_z$.}

We now construct $\tq_s$ for a general $s\geqslant 0$.
For this we use the solution \eqref{solAlphanAs} and the general formula \eqref{puiDcalkYM}.
We obtain\footnote{Recall that $P=\sum_{i=0}^\l p_i$.}
\begin{align}
    \frac{\gYM^2}{-2} \cqAas^u &=\sum_{k=0}^s\int_S\Tr\Big(\tQ_{s-k} \big(\pui\Dcal\big)^k \Aa{s}\Big) \nn\\
    &=\sum_{k=0}^s\int_S\Tr\Bigg( \tQ_{s-k} \sum_{\l=0}^k\sum_{P=k-\l}\pui[p_0]D^{p_0}\bigg(\pui\Big[A,\pui[p_1]D^{p_1}\Big(\pui\Big[A,\pui[p_2]D^{p_2}\Big(\ldots \nn\\
    &\qquad\quad\qquad\ldots \pui\Big[A,\pui[p_{\l-1}]D^{p_{\l-1}}\Big(\pui \Big[A,\pui[p_\l]D^{p_\l}\Aa{s}\Big]_\g\Big)\Big]_\g\ldots \Big)\Big]_\g\Big)\Big]_\g\bigg)\Bigg). \label{intermrenormYM}
\end{align}
Each term in the sum over $k$ behaves similarly.
We thus get (the strategy consists in alternatively integrating by parts over $S$ and using the trace invariance over the adjoint action)\footnote{We use that $\pui[p] \Aa{s} = \frac{u^p}{p!} \Aa{s}$.}${}^,$\footnote{Remember that if there is no opening parenthesis/bracket right after $\pui$, then it acts on everything on its right till the first closing parenthesis/bracket.}
\begin{align}
    \eqref{intermrenormYM} &=\sum_{k=0}^s\int_S\Tr\Bigg(\tQ_{s-k}\bigg(\frac{u^k}{k!}D^k\Aa{s}+\!\!\sum_{P=k-1}\!\!\pui[p_0]D^{p_0}\Big(\pui\Big[A,\pui[p_1]D^{p_1}\Aa{s}\Big]_\g\Big) \nn\\
    &\qquad\qquad+\!\!\sum_{P=k-2} \!\!\pui[p_0]D^{p_0}\Big(\pui\Big[A,\pui[p_1]D^{p_1}\Big(\pui\Big[A,\pui[p_2]D^{p_2}\Aa{s}\Big]_\g\Big)\Big]_\g\Big) \nn\\
    &\qquad\qquad+\ldots+\pui\Big[A,\pui \Big[A,\ldots \pui\Big[A,\Aa{s}\Big]_\g\ldots\Big]_\g\Big]_\g\bigg)\Bigg) \nn\\
    &=\sum_{k=0}^s\int_S\Tr\Bigg((-1)^k \frac{u^k}{k!}D^k\tQ_{s-k}\Aa{s}+\!\!\sum_{P=k-1}\!(-1)^{p_0}D^{p_0}\tQ_{s-k}\Big[\pui[p_0-1]A,\pui[p_1]D^{p_1}\Aa{s}\Big]_\g \nn\\
    &\qquad\qquad+\!\!\sum_{P=k-2}\!(-1)^{p_0}D^{p_0}\tQ_{s-k}\Big[\pui[p_0-1]A,\pui[p_1]D^{p_1}\Big( \pui\Big[A,\pui[p_2]D^{p_2}\Aa{s}\Big]_\g\Big)\Big]_\g \nn\\
    &\qquad\qquad+\ldots+\Big[\tQ_{s-k},\pui\Big(A\Big]_\g\pui\Big[A,\ldots \pui\Big[A,\Aa{s}\Big]_\g\ldots\Big]_\g\Big)\Bigg) \label{intermrenorm2} \\
    &=\sum_{k=0}^s\int_S\Tr\Bigg((-1)^k \frac{u^k}{k!}D^k\tQ_{s-k}\Aa{s}+\!\!\sum_{P=k-1}\!(-1)^{p_0}\Big[D^{p_0}\tQ_{s-k},\pui[p_0-1]\Big(A\Big]_\g \pui[p_1](1)\Big)D^{p_1}\Aa{s}\nn\\
    &\qquad\qquad+\!\!\!\sum_{P=k-2}\!(-1)^{p_0}\Big[D^{p_0}\tQ_{s-k},\pui[p_0-1]\Big(A\Big]_\g\pui[p_1]D^{p_1}\Big( \pui\Big[A,\pui[p_2]D^{p_2}\Aa{s}\Big]_\g\Big)\Big) \nn\\
    &\qquad\qquad+\ldots+\Big[\Big[\tQ_{s-k},\pui\Big(A\Big]_\g,\pui\Big( A\Big]_\g\ldots \pui\Big[A,\Aa{s}\Big]_\g\ldots\Big)\Big)\Bigg) \nn\\
    &=\sum_{k=0}^s\int_S\Tr\Bigg((-1)^k \frac{u^k}{k!}D^k\tQ_{s-k}\Aa{s}+\!\!\sum_{P=k-1}\!(-1)^{k-1}D^{p_1}\Big[D^{p_0}\tQ_{s-k},\pui[p_0-1]\Big(A\Big]_\g \pui[p_1](1)\Big)\Aa{s}\nn\\
    &\qquad\qquad+\!\!\!\sum_{P=k-2}\!(-1)^{p_0+p_1}D^{p_1}\Big[D^{p_0}\tQ_{s-k},\pui[p_0-1]\Big(A\Big]_\g \pui[p_1-1]\Big[A,\pui[p_2]D^{p_2}\Aa{s}\Big]_\g\Big) \nn\\
    &\qquad\qquad+\ldots+\Big[\Big[\Big[ \tQ_{s-k},\pui\Big(A\Big]_\g,\pui\Big( A\Big]_\g,\pui\Big(A\Big]_\g\ldots \pui\Big[A,\Aa{s}\Big]_\g\ldots\Big)\Big)\Big)\Bigg) \nn\\
    &=\sum_{k=0}^s\int_S\Tr\bigg((-1)^k \sum_{\l=0}^k(-1)^\l\!\!\sum_{P=k-\l}\!D^{p_0}\Big[D^{p_1}\Big[D^{p_2}\Big[\ldots \Big[D^{p_\l}\tQ_{s-k},\pui[p_\l-1]\Big(A\Big]_\g,\ldots\Big]_\g,\,\cdot \nn\\
    &\hspace{6cm}\cdot\pui[p_2-1]\Big(A\Big]_\g,\pui[p_1-1]\Big(A\Big]_\g\pui[p_0](1)\Big)\Big)\ldots\Big)\bigg)\Aa{s}. \nn
\end{align}
Hence the renormalized charge aspect for arbitrary spin-weight $s$ takes the form
\begin{empheq}[box=\fbox]{align}
    \tq_s &=\sum_{k=0}^s\sum_{\l=0}^k (-1)^{k+\l}\!\!\sum_{P=k-\l}\!D^{p_0}\Big[D^{p_1}\Big[D^{p_2}\Big[\ldots \Big[D^{p_\l}\tQ_{s-k},\pui[p_\l-1]\Big(A\Big]_\g,\ldots\Big]_\g,\,\cdot \nn\\
    &\hspace{5cm}\cdot\pui[p_2-1]\Big(A\Big]_\g,\pui[p_1-1]\Big(A\Big]_\g\pui[p_0](1)\Big)\Big)\ldots\Big). \label{deftqsYM}
\end{empheq}

\section{Proof of Lemma \hyperref[YMHardAction]{[Hard action]} \label{App:HardActionYM}}

From the generalized Leibniz rule for pseudo-differential calculus \cite{PseudoDiffBakas},
\begin{equation}
    \pui[\beta](fg)=\sum_{n=0}^\infty\frac{(-\beta)_n}{n!}(\pa_u^nf)\pui[(n+\beta)]g, \qquad \beta\in\R, \label{LeibnizRuleGeneral}
\end{equation}
where $(\beta)_n=\beta(\beta-1)\cdots(\beta-n+1)$ is the falling factorial, we get in particular that
\begin{equation}
    \pui[p]\left(\frac{u^{s-p}}{(s-p)!}A\right)=\sum_{n=0}^{s-p}\frac{(-p)_n}{n!} \frac{u^{s-p-n}}{(s-p-n)!}\pui[(n+p)]A.
\end{equation}
From there we can compute \eqref{puiDcalHard} using \eqref{puiDcalkYM}:
\begin{align}
    \hard{\Big(\pa_u\big(\pui\Dcal\big)^{s+1} \Aa{s}\Big)} &=\sum_{p_0+p_1=s} \pui[p_0+1]D^{p_0}\Big(\pui\Big[A,\pui[p_1]D^{p_1}\Aa{s}\Big]_\g\Big) \nn\\
    &=\sum_{p=0}^sD^{p}\Big(\pui[p] \Big[A,\frac{u^{s-p}}{(s-p)!}D^{s-p}\Aa{s}\Big]_\g\Big) \nn\\
    &=\sum_{p=0}^s\sum_{n=0}^{s-p}\frac{(-p)_n}{n!}\frac{u^{s-p-n}}{(s-p-n)!} D^p\big[\pui[(n+p)]A,D^{s-p}\Aa{s}\big]_\g \\
    &=\sum_{k=0}^s\sum_{n=0}^k\frac{(-(k-n))_n}{n!}\frac{u^{s-k}}{(s-k)!} D^{k-n}\big[\pui[k]A,D^{n+s-k}\Aa{s}\big]_\g, \nn
\end{align}
where in the last step we changed variable $n+p\to k$.
Therefore,
\begin{equation}
    \Hd{s}{\Aa{s}}A=\sum_{p=0}^s \frac{u^{s-p}}{(s-p)!}\Htd{p}{D^{s-p}\Aa{s}}A,
\end{equation}
with\footnote{$(-(p-n))_n=(-1)^n(p-1)_n$.}
\begin{equation}
    \Htd{p}{\Aa{p}}A=\sum_{n=0}^p(-1)^{n+1} \frac{(p-1)_n}{n!}D^{p-n}\big[ \pui[p]A,D^n\Aa{p}\big]_\g. \label{tdpHYM}
\end{equation}
When $p=0$, then $\Htd{0}{\Aa0}A=-[A,\Aa0]_\g$, but if $p>0$ then we can simplify \eqref{tdpHYM} much further. 
Indeed, using that $(p-1)_p=0$, we have that
\begin{align}
    \Htd{p}{\Aa{p}}A &=-\sum_{n=0}^{p-1}(-1)^n \frac{(p-1)_n}{n!}\sum_{k=0}^{p-n} \frac{(p-n)_k}{k!}\big[ \pui[p]D^k A,D^{p-k}\Aa{p}\big]_\g \nn\\
    &=-\sum_{k=1}^p\big[\pui[p]D^kA,D^{p-k}\Aa{p}\big]_\g \sum_{n=0}^{p-k}(-)^{n} \frac{(p-1)_n}{n!}\frac{(p-n)_k}{k!}-\sum_{n=0}^{p-1}(-)^n \frac{(p-1)_n}{n!}\big[\pui[p]A,D^p\Aa{p}\big]_\g \nn\\
    &=-\sum_{k=1}^p\big[\pui[p]D^kA,D^{p-k}\Aa{p}\big]_\g\binom{p}{k}{}_2F_1\big(\!-(p-k),1-p;-p;1\big). \label{intermYMHardAction}
\end{align}
To get the first line, we just used the Leibniz rule on $D^{p-n}$.
In the second equality, we extracted the term $k=0$ in order to then be able to commute the sums over $n$ and $k$.
The third line is a rewriting of the binomial coefficients in terms of hypergeometric functions, which will allow us to use the identity (cf. formula 7.3.5.4 in \cite{HyperGmathbook})
\begin{equation}\label{HyperID}
    {}_2F_1(-m,-b;-c;1)=\sum_{n=0}^m(-1)^n \binom{m}{n}\frac{(b)_n}{(c)_n}=\frac{(c-b)_m}{(c)_m}.
\end{equation}
So to get the third line of \eqref{intermYMHardAction}, we used that
\begin{align}
    \sum_{n=0}^{p-k}(-1)^n \frac{(p-1)_n}{n!}\frac{(p-n)_k}{k!} &=\sum_{n=0}^{p-k}(-1)^n \binom{p-k}{n}\frac{(p-1)_n(p-n)!}{(p-k)!k!} \nn\\
    &=\binom{p}{k}\sum_{n=0}^{p-k}(-1)^n \binom{p-k}{n}\frac{(p-1)_n}{(p)_n} \\
    &=\binom{p}{k}{}_2F_1\big(\!-(p-k),1-p;-p;1\big), \nn
\end{align}
together with the binomial identity
\begin{align}
    \sum_{n=0}^{p-1}(-1)^n \binom{p-1}{n}=0.
\end{align}
By \eqref{HyperID}, we infer that for $p\geq 1$,\footnote{In terms of Kronecker $\delta$, $(1)_{p-k}=\delta_{p,k}+\delta_{p-1,k}$.}
\begin{align}
    \Htd{p}{\Aa{p}}A &=-\sum_{k=1}^p\big[\pui[p]D^kA,D^{p-k}\Aa{p}\big]_\g\binom{p}{k}\frac{(1)_{p-k}}{(p)_{p-k}} \nn\\
    &=-\big[\pui[p]D^pA,\Aa{p}\big]_\g- \big[\pui[p]D^{p-1}A,D\Aa{p}\big]_\g=-D\big[\pui[p]D^{p-1}A,\Aa{p}\big]_\g,
\end{align}
which corresponds to the formula \eqref{YMHardAction}.

\section{Hard action: comparison with literature \label{AppYM:comparison}}

Here we show that the equation (50) in \cite{Freidel:2023gue} amounts to \eqref{YMHardAction}.
Indeed, on one hand we get from \cite{Freidel:2023gue} that
\begin{equation} \label{conformalAction3}
    \Hd{s}{\Aa{s}}\hA(\Delta)=i^s\sum_{p=0}^s \frac{(\Delta)_p}{p!}\big[D^p\Aa{s}, 
    D^{s-p}\hA(\Delta-s)\big]_\g,
\end{equation}
where $\hA(\Delta,z,\bz)$\footnote{We keep the dependence in $(z,\bz)$ implicit henceforth.} is the gauge potential in the conformal basis (i.e. the conformal gluon primary operator\footnote{\eqref{ConformalA} is also equal to the Mellin transform in energy of the Fourier transform $\tilde A(\omega)$ of $A(u)$, i.e. 
\begin{equation}
    \hA(\Delta)=\int_0^\infty\rd \omega\,\omega^{\Delta-1}\tilde A(\omega).
\end{equation}}):
\begin{equation} \label{ConformalA}
    \hA(\Delta):=i^\Delta\Gamma(\Delta) \int_\R\rd u\,(u+i\epsilon)^{-\Delta}A(u).
\end{equation}
$\Gamma$ denotes the Gamma function.
On the other hand, we can apply the change to the conformal basis to \eqref{YMHardAction}.
For this we use the following properties---that can be proven by induction:\footnote{The operator $u^{-n} \pui[n]=(\hD+n-1)_n^{-1}$ is the formal inversion of $\pa_u^nu^n=(\hD+n-1)_n$. We prove the latter by induction using that $\pa_u^{n+1}u^{n+1}=\pa_u\big((\pa_u^nu^n)u\big)=(\hD+n)_n\pa_u u=(\hD+n)_{n+1}$, which itself follows from $\pa_u(\hD+\alpha)_n=(\hD+\alpha+1)_n\pa_u$.}
\begin{align} \label{conformalProperties}
    u^\alpha(\hD+n-1)_n^{-1}=(\hD+n-1-\alpha)_n^{-1}u^\alpha,\qquad
    u^{-n} \pui[n] =(\hD+n-1)_n^{-1},
\end{align}
valid for $n\in\N$ and $\alpha\in\R$ and where $\hD:=\pa_u u=1+u\pa_u$.
From \eqref{YMHardAction}, we thus obtain that\footnote{We keep the $i\epsilon$ prescription implicit for shortness. It is understood that it appears all along the calculation, as in \eqref{ConformalA}.
Besides, we also write $\int_\R\equiv \int_\R\rd u$ since the context is clear.}
\begin{align} \label{ConformalAction1}
    \Hd{s}{\Aa{s}}\hA(\Delta)&=-i^\Delta \Gamma(\Delta)\int_\R u^{-\Delta}\left(\frac{u^s}{s!}[A,D^s\Aa{s}]_\g+\sum_{p=1}^s\frac{u^{s-p}}{(s-p)!}D\big[\pui[p]D^{p-1}A,D^{s-p}\Aa{s}\big]_\g\right).
\end{align}
We then use \eqref{conformalProperties} to infer that
\begin{align}
    \int_\R u^{s-\Delta-p}\pui[p]A &=\int_\R u^{s-\Delta}(\hD+p-1)^{-1}_pA \nn\\
    &=\int_\R(\hD+\Delta-s+p-1)^{-1}_p u^{s-\Delta}A \nn\\
    &=\int_\R\frac{\Gamma(\Delta-s)}{\Gamma(\Delta-s+p)}u^{s-\Delta}A \\
    &=\frac{i^{s-\Delta}}{\Gamma(\Delta-s+p)}\hA(\Delta-s), \nn
\end{align}
where in the penultimate line, we used that $\hD$ (and any analytic function of it) integrates to 0 for fields in the Schwartz space.
Therefore, \eqref{ConformalAction1} simplifies to

\begin{align} \label{ConformalAction2}
    \Hd{s}{\Aa{s}}\hA(\Delta)&=\frac{i^s \Gamma(\Delta)}{s!\Gamma(\Delta-s)}\big[D^s\Aa{s}, \hA(\Delta-s)\big]_\g \nn\\*
    &+\sum_{p=1}^s\frac{i^s\Gamma(\Delta)}{(s-p)!\Gamma(\Delta-s+p)}D\big[D^{s-p}\Aa{s},D^{p-1}\hA(\Delta-s)\big]_\g.
\end{align}
Notice that we can recast the sum as follows: We first act with the total $D$ derivative, change $p\to p-1$ in the second sum and then recombine both sums. This gives
\begin{align}
    &\sum_{p=1}^s\frac{i^s\Gamma(\Delta)}{(s-p)!\Gamma(\Delta-s+p)}D\big[D^{s-p}\Aa{s},D^{p-1}\hA(\Delta-s)\big]_\g \nn\\
    =&\sum_{p=1}^s\frac{i^s\Gamma(\Delta)}{(s-p)!\Gamma(\Delta-s+p)}\big[D^{s-p}\Aa{s},D^{p}\hA(\Delta-s)\big]_\g \nn\\
    +&\sum_{p=0}^{s-1}\frac{i^s\Gamma(\Delta)}{(s-p-1)!\Gamma(\Delta-s+p+1)}\big[D^{s-p}\Aa{s},D^p\hA(\Delta-s)\big]_\g \nn\\
    =&\,\frac{i^s\Gamma(\Delta)}{(s-1)!\Gamma(\Delta-s+1)}\big[D^s\Aa{s},\hA(\Delta-s)\big]_\g+i^s\big[\Aa{s},D^s\hA(\Delta-s)\big]_\g \\
    +&\sum_{p=1}^{s-1}\frac{i^s\Gamma (\Delta+1)}{(s-p)!\Gamma(\Delta-s+p+1)}\big[D^{s-p}\Aa{s},D^p\hA(\Delta-s)\big]_\g. \nn
\end{align}
Hence \eqref{ConformalAction2} reduces to
\begin{align}
    \Hd{s}{\Aa{s}}\hA(\Delta)&=\sum_{p=0}^s \frac{i^s\Gamma (\Delta+1)}{(s-p)!\Gamma(\Delta-s+p+1)}\big[D^{s-p}\Aa{s},D^p\hA(\Delta-s)\big]_\g \nn\\
    &=i^s\sum_{p=0}^s\frac{\Gamma(\Delta+1)}{p!\Gamma(\Delta-p+1)}\big[D^p\Aa{s},D^{s-p}\hA(\Delta-s)\big]_\g,
\end{align}
which is indeed \eqref{conformalAction3}.

\newpage
\addcontentsline{toc}{section}{References}

\bibliographystyle{JHEP}
\bibliography{source_NoetherYM}

\end{document}